# High-Breakdown Robust Multivariate Methods

**Mia Hubert, Peter J. Rousseeuw and Stefan Van Aelst**

*Abstract.* When applying a statistical method in practice it often occurs that some observations deviate from the usual assumptions. However, many classical methods are sensitive to outliers. The goal of robust statistics is to develop methods that are robust against the possibility that one or several unannounced outliers may occur anywhere in the data. These methods then allow to detect outlying observations by their residuals from a robust fit. We focus on high-breakdown methods, which can deal with a substantial fraction of outliers in the data. We give an overview of recent high-breakdown robust methods for multivariate settings such as covariance estimation, multiple and multivariate regression, discriminant analysis, principal components and multivariate calibration.

## 1. INTRODUCTION

Many multivariate datasets contain outliers, that is, data points that deviate from the usual assumptions and/or from the pattern suggested by the majority of the data. Outliers are more likely to occur in datasets with many observations and/or variables, and often they do not show up by simple visual inspection.

The usual multivariate analysis techniques (e.g., principal components, discriminant analysis and multivariate regression) are based on empirical means, covariance and correlation matrices, and least squares fitting. All of these can be strongly affected by even a few outliers. When the data contain nasty outliers, typically two things happen:

- the multivariate estimates differ substantially from the "right" answer, defined here as the estimates we would have obtained without the outliers;
- the resulting fitted model does not allow to detect the outliers by means of their residuals, Mahalanobis distances or the widely used "leave-one-out" diagnostics.

The first consequence is fairly well known (although the size of the effect is often underestimated). Unfortunately, the second consequence is less well known, and when stated many people find it hard to believe or paradoxical. Common intuition says that outliers must "stick out" from the classical fitted model, and indeed some of them may do so. But the most harmful types of outliers, especially if there are several of them, may affect the estimated model so much "in their direction" that they are now well-fitted by it.

*Mia Hubert is Professor, University Center for Statistics and Department of Mathematics, Katholieke Universiteit Leuven, Celestijnenlaan 200 B, B-3001 Leuven, Belgium (e-mail: mia.hubert@wis.kuleuven.be). Peter J. Rousseeuw is Professor, Department of Mathematics and Computer Science, University of Antwerp, Middelheimlaan 1, B-2020 Antwerp, Belgium (e-mail: peter.rousseeuw@ua.ac.be). Stefan Van Aelst is Professor, Department of Applied Mathematics and Computer Science, Ghent University, Krijgslaan 281 S9, B-9000 Ghent, Belgium (e-mail: stefan.vanaelst@UGent.be).*

Once this effect is understood, one sees that the following two problems are essentially equivalent:

- Robust estimation: find a "robust" fit, which is similar to the fit we would have found without the outliers.
- Outlier detection: find all the outliers that matter.

Indeed, a solution to the first problem allows us to identify the outliers by their residuals, and so on, from the robust fit. Conversely, a solution to the second problem allows us to remove or downweight the outliers followed by a classical fit, which yields a robust result.

Our research focuses on the first problem, and uses its results to answer the second. We prefer this approach over the opposite direction because from a combinatorial viewpoint it is more feasible to search for *sufficiently many* "good" data points than to find *all* the "bad" data points.

It turns out that most of the currently available highly robust multivariate estimators are difficult to compute, which makes them unsuitable for the analysis of large and/or high-dimensional datasets. Among the few exceptions is the minimum covariance determinant estimator (MCD) of Rousseeuw (1984, 1985). The MCD is a highly robust estimator of multivariate location and scatter, that can be computed efficiently with the FAST-MCD algorithm of Rousseeuw and Van Driessen (1999).

Section 2 concentrates on robust estimation of location and scatter. We first describe the MCD estimator and discuss its main properties. Alternatives for the MCD are explained briefly with relevant pointers to the literature for more details. Section 3 does the same for robust regression and mainly focuses on the least trimmed squares (LTS) estimator (Rousseeuw, 1984), which is an analog of MCD for multiple regression. Since estimating the covariance matrix is the cornerstone of many multivariate statistical methods, robust scatter estimators have also been used to develop robust and computationally efficient multivariate techniques. The paper then goes on to describe robust methods for multivariate regression (Section 4), classification (Section 5), principal component analysis (Section 6), principal component regression (Section 7), partial least squares regression (Section 8) and other settings (Section 9). Section 10 concludes with pointers to available software for the described techniques.

## 2. MULTIVARIATE LOCATION AND SCATTER

### 2.1 The Need for Robustness

In the multivariate location and scatter setting we assume that the data are stored in an $n \times p$ data matrix $\mathbf{X} = (\mathbf{x}_1, \ldots, \mathbf{x}_n)'$ with $\mathbf{x}_i = (x_{i1}, \ldots, x_{ip})'$ the $i$th observation. Hence $n$ stands for the number of objects and $p$ for the number of variables.

To illustrate the effect of outliers we consider the following engineering problem, taken from Rousseeuw and Van Driessen (1999). Philips Mecoma (The Netherlands) produces diaphragm parts for television sets. These are thin metal plates, molded by a press. When starting a new production line, $p = 9$ characteristics were measured for $n = 677$ parts. The aim is to gain insight in the production process and to find out whether abnormalities have occurred. A classical approach is to compute the Mahalanobis distance

$$\text{(1)} \qquad \text{MD}(\mathbf{x}_i) = \sqrt{(\mathbf{x}_i - \hat{\boldsymbol{\mu}}_0)' \hat{\boldsymbol{\Sigma}}_0^{-1} (\mathbf{x}_i - \hat{\boldsymbol{\mu}}_0)}$$

of each measurement $\mathbf{x}_i$. Here $\hat{\boldsymbol{\mu}}_0$ is the arithmetic mean and $\hat{\boldsymbol{\Sigma}}_0$ is the classical covariance matrix. The distance $\text{MD}(\mathbf{x}_i)$ should tell us how far away $\mathbf{x}_i$ is from the center of the cloud, relative to the size of the cloud.

In Figure 1 we plotted the classical Mahalanobis distance versus the index $i$, which corresponds to the production sequence. The horizontal line is at the usual cutoff value $\sqrt{\chi^2_{9,0.975}} = 4.36$. Figure 1 suggests that most observations are consistent with the classical assumption that the data come from a multivariate normal distribution, except for a few isolated outliers. This should not surprise us, even in the first experimental run of a new production line, because the Mahalanobis distances are known to suffer from the *masking effect*. That is, even if there were a group of outliers (here, deformed diaphragm parts), they would affect $\hat{\boldsymbol{\mu}}_0$ and $\hat{\boldsymbol{\Sigma}}_0$ in such a way that they get small Mahalanobis distances $\text{MD}(\mathbf{x}_i)$ and thus become invisible in Figure 1. To get a reliable analysis of these data we need robust estimators $\hat{\boldsymbol{\mu}}$ and $\hat{\boldsymbol{\Sigma}}$ that can resist possible outliers. For this purpose we will use the MCD estimates described below.

### 2.2 Description of the MCD

The MCD method looks for the $h$ observations (out of $n$) whose classical covariance matrix has the lowest possible determinant. The MCD estimate of location is then the average of these $h$ points, whereas the MCD estimate of scatter is a multiple of their covariance matrix. The MCD location and scatter estimates are affine equivariant, which means that they behave properly under affine transformations of the data. That is, for an $n \times p$ dataset $\mathbf{X}$ the MCD estimates $(\hat{\boldsymbol{\mu}}, \hat{\boldsymbol{\Sigma}})$ satisfy

$$\hat{\boldsymbol{\mu}}(\mathbf{XA} + \mathbf{1}_n \mathbf{v}') = \hat{\boldsymbol{\mu}}(\mathbf{X})\mathbf{A} + \mathbf{v}, \tag{2}$$

$$\hat{\boldsymbol{\Sigma}}(\mathbf{XA} + \mathbf{1}_n \mathbf{v}') = \mathbf{A}'\hat{\boldsymbol{\Sigma}}(\mathbf{X})\mathbf{A}, \tag{3}$$

for all $p \times 1$ vectors $\mathbf{v}$ and all nonsingular $p \times p$ matrices $\mathbf{A}$. The vector $\mathbf{1}_n$ is $(1, 1, \ldots, 1)'$ with $n$ elements. Affine equivariance is a natural property of the underlying model and makes the analysis independent of the measurement scales of the variables as well as translations or rotations of the data.

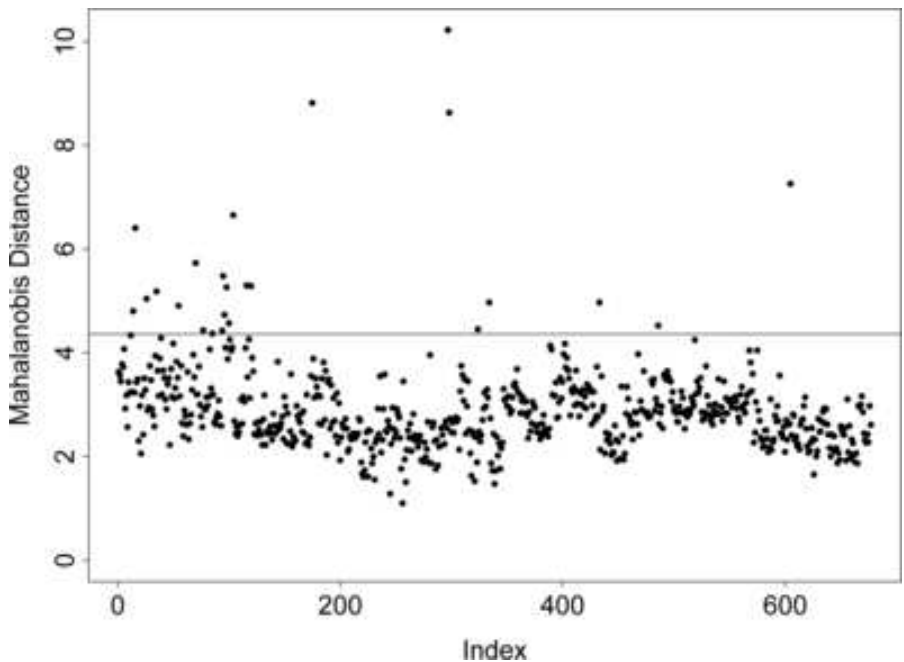

Fig. 1. *Mahalanobis distances of the Philips data.*

A useful measure of robustness is the *finite-sample breakdown value* (Donoho and Huber, 1983). The breakdown value $\varepsilon_n^*(\hat{\boldsymbol{\mu}}, \mathbf{X})$ of an estimator $\hat{\boldsymbol{\mu}}$ at the dataset $\mathbf{X}$ is the smallest amount of contamination that can have an arbitrarily large effect on $\hat{\boldsymbol{\mu}}$. Consider all possible contaminated datasets $\tilde{\mathbf{X}}$ obtained by replacing *any* $m$ of the original observations by *arbitrary* points. Then the breakdown value of a location estimator $\hat{\boldsymbol{\mu}}$ is the smallest fraction $m/n$ of outliers that can take the estimate over all bounds:

$$\begin{aligned} &\varepsilon_n^*(\hat{\boldsymbol{\mu}}, \mathbf{X}) \\ &:= \min_m \left\{ \frac{m}{n}; \sup_{\tilde{\mathbf{X}}} \|\hat{\boldsymbol{\mu}}(\tilde{\mathbf{X}}) - \hat{\boldsymbol{\mu}}(\mathbf{X})\| = \infty \right\}. \end{aligned} \tag{4}$$

For many estimators $\varepsilon_n^*(\hat{\boldsymbol{\mu}}, \mathbf{X})$ varies only slightly with $\mathbf{X}$ and $n$, so that we can denote its limiting value (for $n \to \infty$) by $\varepsilon^*(\hat{\boldsymbol{\mu}})$. Similarly, the breakdown value of a covariance matrix estimator $\hat{\boldsymbol{\Sigma}}$ is defined as the smallest fraction of outliers that can take either the largest eigenvalue $\lambda_1(\hat{\boldsymbol{\Sigma}})$ to infinity or the smallest eigenvalue $\lambda_p(\hat{\boldsymbol{\Sigma}})$ to zero. The MCD estimates $(\hat{\boldsymbol{\mu}}, \hat{\boldsymbol{\Sigma}})$ of multivariate location and scatter have breakdown value $\varepsilon_n^*(\hat{\boldsymbol{\mu}}) = \varepsilon_n^*(\hat{\boldsymbol{\Sigma}}) \approx (n - h)/n$. The MCD has its highest possible breakdown value ($\varepsilon^* = 50\%$) when $h = [(n + p + 1)/2]$ (see Lopuhaä and Rousseeuw, 1991). Note that no affine equivariant estimator can have a breakdown value above 50%. For a recent discussion of the importance of equivariance in breakdown considerations, see Davies and Gather (2005).

An efficient algorithm to compute the MCD is the FAST-MCD algorithm explained in Appendix A.1. By default FAST-MCD computes a one-step weighted estimate given by

$$\hat{\boldsymbol{\mu}}_1 = \left( \sum_{i=1}^n w_i \mathbf{x}_i \right) \Big/ \left( \sum_{i=1}^n w_i \right), \tag{5}$$

$$\hat{\boldsymbol{\Sigma}}_1 = d_{h,n} \left( \sum_{i=1}^n w_i (\mathbf{x}_i - \hat{\boldsymbol{\mu}}_1)(\mathbf{x}_i - \hat{\boldsymbol{\mu}}_1)' \right) \tag{6}$$

$$\cdot \left( \sum_{i=1}^n w_i \right)^{-1},$$

where

$$w_i = \begin{cases} 1, & \text{if } d_{(\hat{\boldsymbol{\mu}}_{\text{MCD}}, \hat{\boldsymbol{\Sigma}}_{\text{MCD}})}(i) \leq \sqrt{\chi^2_{p,0.975}}, \\ 0, & \text{otherwise}, \end{cases}$$

with $\hat{\boldsymbol{\mu}}_{\text{MCD}}$ and $\hat{\boldsymbol{\Sigma}}_{\text{MCD}}$ the raw MCD estimates. The number $d_{h,n}$ in (6) is a correction factor (Pison, Van Aelst and Willems, 2002) to obtain unbiased and consistent estimates when the data come from a multivariate normal distribution.

This one-step weighted estimator has the same breakdown value as the initial MCD estimator but a much better statistical efficiency. In practice we often do not need the maximal breakdown value. For example, Hampel et al. (1986, pages 27–28) write that 10% of outliers is quite common. We typically use $h = 0.75n$ so that $\varepsilon^* = 25\%$, which is sufficiently robust for most applications and has a high statistical efficiency. For example, with $h = 0.75n$ the asymptotic efficiencies of the weighted MCD location and scatter estimators in 10 dimensions are 94% and 88%, respectively (Croux and Haesbroeck, 1999).

### 2.3 Examples

Let us now reanalyze the Philips data. For each observation $\mathbf{x}_i$ we now compute the robust distance (Rousseeuw and Leroy, 1987) given by

$$(7) \qquad \text{RD}(\mathbf{x}_i) = \sqrt{(\mathbf{x}_i - \hat{\boldsymbol{\mu}})'\hat{\boldsymbol{\Sigma}}^{-1}(\mathbf{x}_i - \hat{\boldsymbol{\mu}})},$$

where $(\hat{\boldsymbol{\mu}}, \hat{\boldsymbol{\Sigma}})$ are the MCD location and scatter estimates. Recall that the Mahalanobis distances in Figure 1 indicated no groups of outliers. On the other hand, the robust distances $\text{RD}(\mathbf{x}_i)$ in Figure 2 show a strongly deviating group of outliers, ranging from index 491 to index 565. Something happened in the production process, which was not visible from the classical Mahalanobis distances due to the masking effect. Furthermore, Figure 2 also shows a remarkable change after the first 100 measurements. Both phenomena were investigated and interpreted by the engineers at Philips.

The second dataset came from a group of Cal Tech astronomers working on the Digitized Palomar Sky Survey (see Odewahn et al., 1998). They made a survey of celestial objects (light sources) by recording nine characteristics (such as magnitude, area, image moments) in each of three bands: blue, red and near-infrared. The database contains measurements for 27 variables on 137,256 celestial objects. Based on exploratory data analysis Rousseeuw and

Van Driessen (1999) selected six of the variables (two from each band). The classical Mahalanobis distances revealed a set of outliers which turned out to be objects for which at least one measurement fell outside its physically possible range. Therefore, the data was cleaned by removing all objects with physically impossible measurements, leading to a cleaned dataset of size 132,402. The Mahalanobis distances of the cleaned data are shown in Figure 3(a).

This plot (and a Q–Q plot) suggests that the distances approximately come from the $\sqrt{\chi^2_6}$ distribution, as would be the case if the data came from a homogeneous population. Figure 3(b) shows the robust distances computed with the FAST-MCD algorithm. In contrast to the innocent-looking Mahalanobis distances, these robust distances reveal the presence of two groups. There is a majority with $\text{RD}(\mathbf{x}_i) \leq \sqrt{\chi^2_{6,0.975}}$ and a group with $\text{RD}(\mathbf{x}_i)$ between 8 and 16. Based on these results the astronomers noted that the lower group are mainly stars while the upper group are mainly galaxies.

### 2.4 Other robust estimators of multivariate location and scatter

The breakdown point is not the only important robustness measure. Another key concept is the influence function, which measures the effect on an estimator of adding a small mass at a specific point. (See Hampel et al., 1986 for details.) Robust estimators ideally have a bounded influence function, which means that a small contamination at any point can only have a small effect on the estimator. M-estimators (Maronna, 1976; Huber, 1981) were the first class of bounded influence estimators for multivariate location and scatter. Also the MCD and other estimators mentioned below have a bounded influence function. The first high-breakdown location and scatter estimator was proposed by Stahel (1981) and Donoho (1982). The Stahel–Donoho estimates are a weighted mean and covariance, like (5)–(6), where the weight $w_i$ of an observation $\mathbf{x}_i$ depends on its *outlyingness*, given by

$$u_i = \sup_{\|\mathbf{v}\|=1} \frac{|\mathbf{x}_i'\mathbf{v} - \text{med}_j(\mathbf{x}_j'\mathbf{v})|}{\text{mad}_j(\mathbf{x}_j'\mathbf{v})}.$$

The estimator has good robustness properties but is computationally very intensive, which limits its use (Tyler, 1994; Maronna and Yohai, 1995). The Stahel–Donoho estimator measures the outlyingness by looking at all univariate projections of the data and as such is related to projection pursuit methods as studied in Friedman and Tukey (1974), Huber (1985) and Croux and Ruiz-Gazen (2005). Another highly robust estimator of location and scatter based on projections has been proposed by Maronna, Stahel and Yohai (1992).

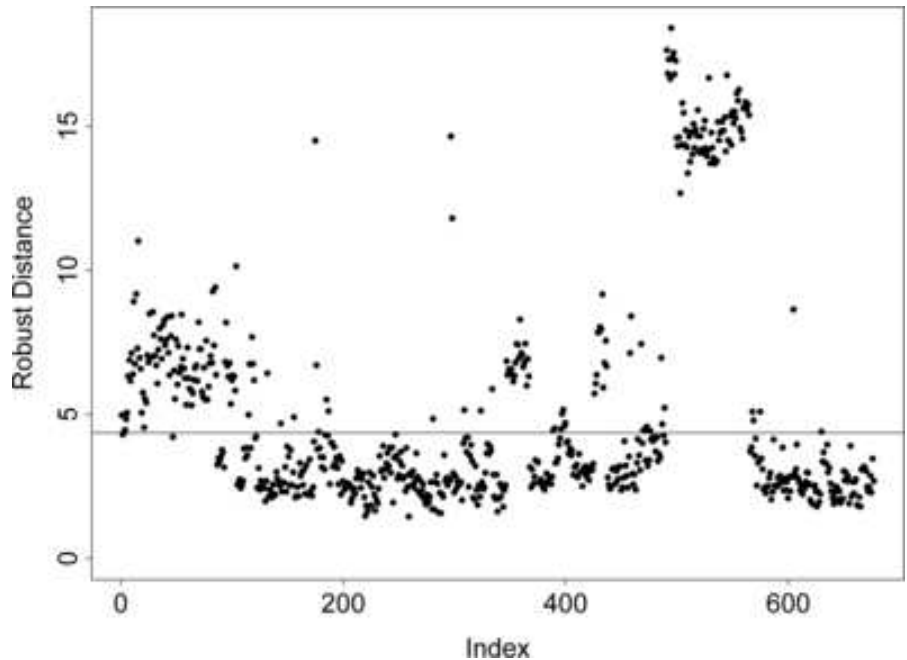

FIG. 2. *Robust distances of the Philips data.*

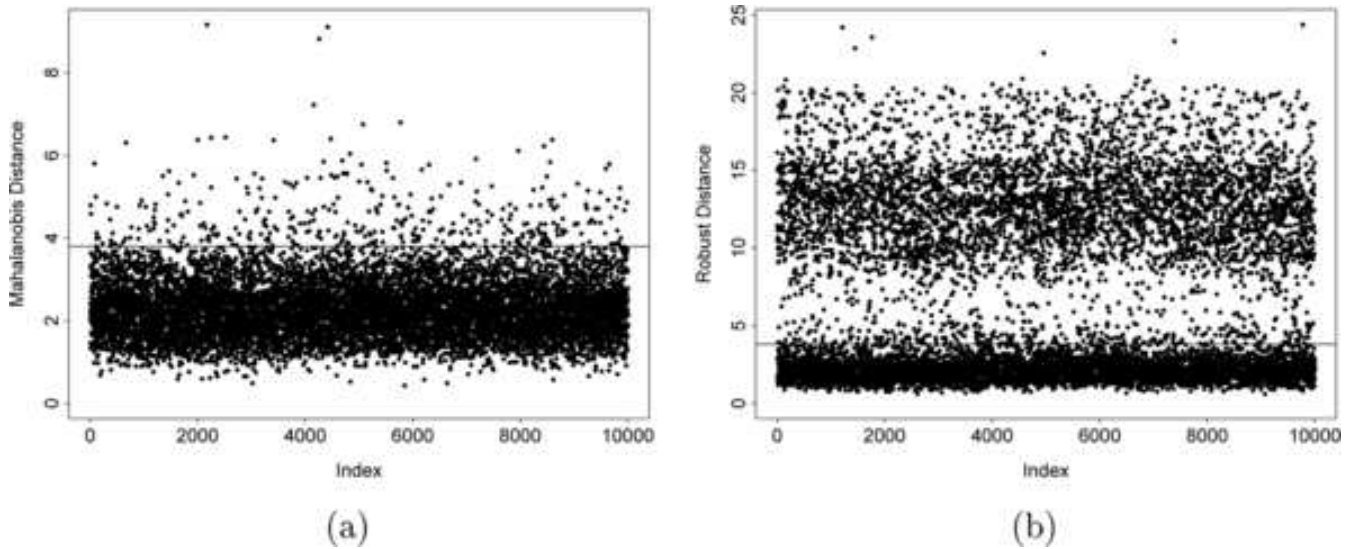

FIG. 3. *Cleaned digitized Palomar data:* (a) *Mahalanobis distances;* (b) *robust distances.*

Together with the MCD, Rousseeuw (1984, 1985) also introduced the minimum volume ellipsoid (MVE) estimator which looks for the minimal volume el-lipsoid covering at least half the data points. However, the MVE has efficiency zero due to its low rate of convergence. Rigorous asymptotic results for the MCD and the MVE are given by Butler, Davies and Jhun (1993) and Davies (1992a). To improve the finite-sample efficiency of MVE and MCD a one-step weighted estimator (5)–(6) can be computed. The breakdown value and asymptotic properties of one-step weighted estimators have been obtained by

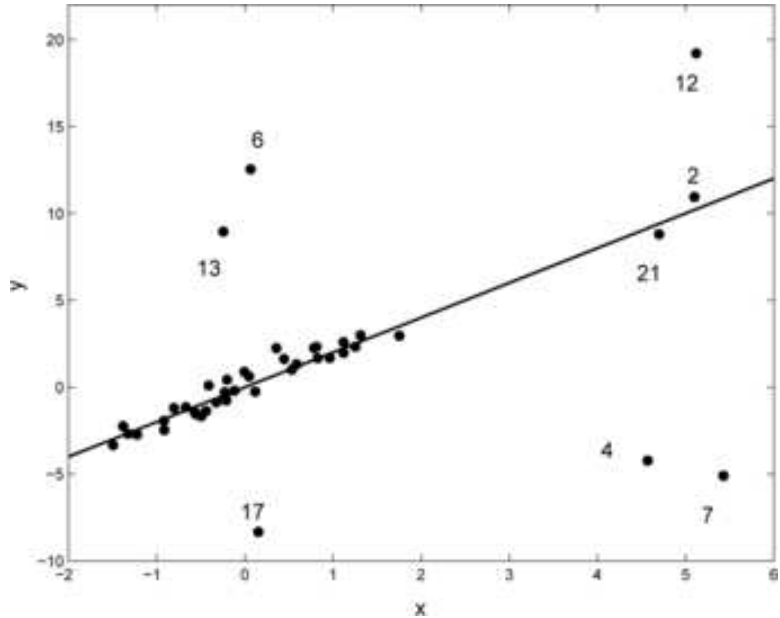

Fig. 4. *Simple regression data with different types of outliers.*

Lopuhaä and Rousseeuw ([1991](#)) and Lopuhaä ([1999](#)). Alternatively, a one-step M-estimator starting from MVE or MCD can be computed as proposed by Davies ([1992b](#)).

Another approach to improve the efficiency of MVE or MCD is to use a smoother objective function. An important class of robust estimators of multivariate location and scatter are S-estimators (Rousseeuw and Leroy, [1987](#); Davies, [1987](#)), defined as the solution $(\hat{\boldsymbol{\mu}}, \hat{\boldsymbol{\Sigma}})$ which minimizes $\det(\boldsymbol{\Sigma})$ under the constraint

$$(8) \qquad \frac{1}{n} \sum_{i=1}^{n} \rho\left(\sqrt{(\mathbf{x}_i - \boldsymbol{\mu})' \boldsymbol{\Sigma}^{-1} (\mathbf{x}_i - \boldsymbol{\mu})}\right) \le b$$

over all vectors $\boldsymbol{\mu}$ and all $p \times p$ positive definite symmetric matrices $\boldsymbol{\Sigma}$. Setting $b = E_F[\rho(\|\mathbf{X}\|)]$ assures consistency at the model distribution $F$. The function $\rho$ is chosen by the statistician and is often taken to be Tukey's biweight $\rho$-function

$$(9) \qquad \rho(x) = \begin{cases} \dfrac{x^2}{2} - \dfrac{x^4}{2c^2} + \dfrac{x^6}{6c^4}, & \text{if } |x| \le c, \\ \dfrac{c^2}{6}, & \text{if } |x| \ge c. \end{cases}$$

The constant $c$ determines the breakdown value which is given by $\varepsilon^* = 6b/c^2$. The properties of S-estimators have been investigated by Lopuhaä ([1989](#)). Related classes include CM-estimators (Kent and Tyler, [1996](#)), MM-estimators (Tatsuoka and Tyler, [2000](#)) and $\tau$-estimators Lopuhaä ([1991](#)). Positive-breakdown estimators of location and scatter can also be used to construct formal outlier identification rules; see, for example, Becker and Gather ([1999](#)).

To extend the notion of ranking to higher dimensions, Tukey introduced the *halfspace depth*. Depth-based estimators have been proposed and studied by Donoho and Gasko ([1992](#)), Rousseeuw, Ruts and Tukey ([1999a](#)), Liu, Parelius and Singh ([1999](#)), Zuo and Serfling ([2000a](#), [2000b](#)) and Zuo, Cui and He ([2004](#)).

Robust estimation and outlier detection in higher dimensions has been studied by Rocke ([1996](#)) and Rocke and Woodruff ([1996](#)). For very high-dimensional data, Maronna and Zamar ([2002](#)) and Alqallaf et al. ([2002](#)) proposed computationally efficient robust estimators of multivariate location and covariance that are not affine equivariant any more. Chen and Victoria-Feser ([2002](#)) address robust covariance matrix estimation with missing data.

## 3. MULTIPLE REGRESSION

### 3.1 Motivation

The multiple regression model assumes that also a response variable $y$ is measured, which can be explained as an affine combination of the $x$-variables. More precisely, the model says that for all observations $(\mathbf{x}_i, y_i)$ with $i = 1, \ldots, n$ it holds that

$$(10) \qquad y_i = \theta_1 x_{i1} + \cdots + \theta_p x_{ip} + \theta_{p+1} + \varepsilon_i, \\ i = 1, \ldots, n,$$

where the errors $\epsilon_i$ are assumed to be i.i.d. with zero mean and constant variance $\sigma^2$. The vector $\boldsymbol{\beta} = (\theta_1, \ldots, \theta_p)'$ is called the slope, and $\alpha = \theta_{p+1}$ the intercept. Denote $\mathbf{x}_i = (x_{i1}, \ldots, x_{ip})'$ and $\boldsymbol{\theta} = (\boldsymbol{\beta}', \alpha)' = (\theta_1, \ldots, \theta_p, \theta_{p+1})'$.

The classical least squares method to estimate $\boldsymbol{\theta}$ and $\sigma$ is extremely sensitive to regression outliers, that is, observations that do not obey the linear pattern formed by the majority of the data. In regression we can distinguish between different types of points. This is illustrated in Figure 4 for simple regression. *Leverage points* are observations $(\mathbf{x}_i, y_i)$ whose $\mathbf{x}_i$ are outlying; that is, $\mathbf{x}_i$ deviates from the majority in $x$-space. We call such an observation $(\mathbf{x}_i, y_i)$ a good leverage point if $(\mathbf{x}_i, y_i)$ follows the linear pattern of the majority, such as points 2 and 21. If, on the other hand, $(\mathbf{x}_i, y_i)$ does not follow this linear pattern, we call it a bad leverage point, like 4, 7 and 12. An observation whose $\mathbf{x}_i$ belongs to the majority in $x$-space but where $(\mathbf{x}_i, y_i)$ deviates from the linear pattern is called a vertical outlier, like the points 6, 13 and 17. A regression dataset can thus have up to four types of points: regular observations, vertical outliers, good leverage points and bad leverage points. Leverage points attract the least squares solution toward them, so bad leverage points are often not apparent in a classical regression analysis.

In low dimensions, as in this example, visual inspection can be used to detect outliers and leverage points, but in higher dimensions this is not an option anymore. Therefore, we need robust and computationally efficient estimators that yield a reliable analysis of regression data. We consider the least trimmed squares estimator (LTS) proposed by Rousseeuw (1984) for this purpose.

For a dataset $\mathbf{Z} = \{(\mathbf{x}_i, y_i); i = 1, \ldots, n\}$ and for any $\boldsymbol{\theta}$ denote the corresponding residuals by $r_i = r_i(\boldsymbol{\theta}) = y_i - \boldsymbol{\beta}'\mathbf{x}_i - \alpha = y_i - \boldsymbol{\theta}'\mathbf{u}_i$ with $\mathbf{u}_i = (\mathbf{x}_i', 1)'$. Then the LTS estimator is defined as the $\hat{\boldsymbol{\theta}}$ which minimizes

$$(11) \qquad \sum_{i=1}^{h} (r^2)_{i:n},$$

where $(r^2)_{1:n} \leq (r^2)_{2:n} \leq \cdots \leq (r^2)_{n:n}$ are the ordered squared residuals (note that the residuals are first squared and then ordered). This is equivalent to finding the $h$-subset with smallest least squares objective function, which resembles the definition of the MCD. The LTS estimate is then the least squares fit to these $h$ points. The LTS estimates are regression, scale and affine equivariant. That is, for any $\mathbf{X} = (\mathbf{x}_1, \ldots, \mathbf{x}_n)'$ and $\mathbf{y} = (y_1, \ldots, y_n)'$ it holds that

$$\hat{\boldsymbol{\theta}}(\mathbf{X}, \mathbf{y} + \mathbf{X}\mathbf{v} + \mathbf{1}_n c) = \hat{\boldsymbol{\theta}}(\mathbf{X}, \mathbf{y}) + (\mathbf{v}', c)'$$

$$(12) \qquad \hat{\boldsymbol{\theta}}(\mathbf{X}, c\mathbf{y}) = c\hat{\boldsymbol{\theta}}(\mathbf{X}, \mathbf{y}),$$

$$\hat{\boldsymbol{\theta}}(\mathbf{X}\mathbf{A}' + \mathbf{1}_n \mathbf{v}', \mathbf{y}) = (\hat{\boldsymbol{\beta}}'(\mathbf{X}, \mathbf{y})\mathbf{A}^{-1}, \alpha(\mathbf{X}, \mathbf{y}) - \hat{\boldsymbol{\beta}}'(\mathbf{X}, \mathbf{y})\mathbf{A}^{-1}\mathbf{v})',$$

for any vector $\mathbf{v}$, any constant $c$ and any nonsingular $p \times p$ matrix $\mathbf{A}$.

The breakdown value of a regression estimator $\hat{\boldsymbol{\theta}}$ at a dataset $\mathbf{Z}$ is the smallest fraction of outliers that can have an arbitrarily large effect on $\hat{\boldsymbol{\theta}}$. Formally, it is defined by (4) where $\mathbf{X}$ is replaced by $(\mathbf{X}, \mathbf{y})$. For $h = [(n + p + 1)/2]$ the LTS breakdown value equals $\varepsilon^*(\text{LTS}) \approx 50\%$, whereas for larger $h$ we have that $\varepsilon_n^*(\text{LTS}) \approx (n - h)/n$. The usual choice $h \approx 0.75\,n$ yields $\varepsilon^*(\text{LTS}) = 25\%$.

When using LTS regression, the standard deviation of the errors can be estimated by

$$(13) \qquad \hat{\sigma} = c_{h,n} \sqrt{\frac{1}{h} \sum_{i=1}^{h} (r^2)_{i:n}},$$

where $r_i$ are the residuals from the LTS fit, and $c_{h,n}$ makes $\hat{\sigma}$ consistent and unbiased at Gaussian error distributions (Pison, Van Aelst and Willems, 2002). Note that the LTS scale estimator $\hat{\sigma}$ is itself highly robust. Therefore, we can identify regression outliers by their standardized LTS residuals $r_i/\hat{\sigma}$.

To compute the LTS in an efficient way, Rousseeuw and Van Driessen (2006) developed the FAST-LTS algorithm outlined in Appendix A.2. Similarly to the FAST-MCD algorithm, FAST-LTS returns weighted least squares estimates, given by

$$(14) \qquad \hat{\boldsymbol{\theta}}_1 = \left( \sum_{i=1}^{n} w_i \mathbf{u}_i \mathbf{u}_i' \right)^{-1} \left( \sum_{i=1}^{n} w_i \mathbf{u}_i y_i \right),$$

$$(15) \qquad \hat{\sigma}_1 = d_{h,n} \sqrt{\frac{\sum_{i=1}^{n} w_i r_i(\hat{\boldsymbol{\theta}}_1)^2}{\sum_{i=1}^{n} w_i}},$$

where $\mathbf{u}_i = (\mathbf{x}_i', 1)'$. The weights are

$$w_i = \begin{cases} 1, & \text{if } |r_i(\hat{\boldsymbol{\theta}}_{\text{LTS}})/\hat{\sigma}_{\text{LTS}}| \leq \sqrt{\chi^2_{1,0.975}}, \\ 0, & \text{otherwise.} \end{cases}$$

where $\hat{\boldsymbol{\theta}}_{\text{LTS}}$ and $\hat{\sigma}_{\text{LTS}}$ are the raw LTS estimates. As before, $d_{h,n}$ is a finite-sample correction factor.

These weighted estimates have the same breakdown value as the initial LTS estimates and a much better statistical efficiency. Moreover, from the weighted least squares estimates all the usual inferential output such as $t$-statistics, $F$-statistics an $R^2$ statistic and the corresponding $p$-values can be obtained (Rousseeuw and Leroy, 1987). These $p$-values are approximate since they assume that the data with $w_i = 1$ come from the model (10) whereas the data with $w_i = 0$ do not, and we usually do not know whether that is true.

In Figure 4 we see that the LTS line obtained by FAST-LTS yields a robust fit that is not attracted by the leverage points on the right-hand side, and hence follows the pattern of the majority of the data. Of course, the LTS method is most useful when there are several $x$-variables.

To detect leverage points in higher dimensions we must detect outlying $\mathbf{x}_i$ in $x$-space. For this purpose we will use the robust distances $RD_i$ based on the one-step weighted MCD of the previous section. On the other hand, we can see whether a point $(\mathbf{x}_i, y_i)$ lies near the majority pattern by looking at its standardized LTS residual $r_i/\hat{\sigma}$. Rousseeuw and van Zomeren (1990) proposed an outlier map which plots robust residuals $r_i/\hat{\sigma}$ versus robust distances $RD(\mathbf{x}_i)$, and indicates the corresponding cutoffs by horizontal and vertical lines. It automatically classifies the observations into the four types of data points that can occur in a regression dataset. Figure 5 is the outlier map of the data in Figure 4.

To illustrate this plot, we again consider the database of the Digitized Palomar Sky Survey. Following Rousseeuw and Van Driessen (2006), we now use the subset of 56,744 stars (not galaxies) for which all the characteristics in the blue color (the F band) are available. The response variable MaperF is regressed against the other eight characteristics of the color band F. These characteristics describe the size of a light source and the shape of the spatial brightness distribution in a source. Figure 6(a) plots the standardized LS residuals versus the classical Mahalanobis distances. Some isolated outliers in the $y$-direction as well as in $x$-space were not plotted to get a better view of the majority of the data. Observations for which the standardized absolute LS residual exceeds the cutoff $\sqrt{\chi^2_{1,0.975}}$ are considered to be regression outliers, whereas the other observations are thought to obey the linear model. Similarly, observations for which $MD(\mathbf{x}_i)$ exceeds the cutoff

$\sqrt{\chi^2_{8,0.975}}$ are considered to be leverage points. Figure 6(a) shows that most data points lie between the horizontal cutoffs at $\pm\sqrt{\chi^2_{1,0.975}}$ which suggests that most data follow the same linear trend. On the other hand, the outlier map based on LTS residuals and robust distances $RD(\mathbf{x}_i)$ shown in Figure 6(b) tells a different story. This plot reveals a rather large group of observations with large robust residuals and large robust distances. Hence, these observations are bad leverage points. This group turned out to be giant stars, which are known to behave differently from other stars.

## 3.2 Other robust regression methods

The development of robust regression often paralleled that of robust estimators of multivariate location and scatter, and in fact more attention has been dedicated to the regression setting. Robust regression also started with M-estimators (Huber, 1973, 1981), later followed by R-estimators (Jurecková, 1971) and L-estimators (Koenker and Portnoy, 1987) that all have breakdown value zero because of their vulnerability to bad leverage points.

The next step was the development of generalized M-estimators (GM-estimators) that bound the influence of outlying $\mathbf{x}_i$ by giving them a small weight (see, e.g., Krasker and Welsch, 1982; Maronna and Yohai, 1981). Therefore, GM-estimators are often called *bounded influence methods*, and they are more stable than M-, L- or R-estimators. See Hampel et al. (1986, Chapter 6) for an overview. Unfortunately, the breakdown value of GM-estimators with a monotone score function still goes down to zero for increasing $p$ (Maronna, Burtos and Yohai, 1979). GM-estimators with a redescending score function can have a dimension-independent positive breakdown value (see He, Simpson and Wang, 2000). Note that for a small fraction of outliers in the data GM-estimators are robust, and they are computationally fast. For a discussion of the differences between bounded-influence estimators and high-breakdown methods see the recent book by Maronna, Martin and Yohai (2006).

The first high-breakdown regression methods were least median of squares (LMS), LTS and the repeated median. The origins of LMS go back to Tukey (Andrews et al., 1972), who proposed a univariate estimator based on the shortest half of the sample and called it the *shorth*. Hampel (1975, page 380) modified and generalized it to regression and stated that the resulting estimator has a 50% breakdown value. He called it the *shordth* and considered it of special mathematical interest. Later, Rousseeuw (1984) provided theory, algorithms and programs for this estimator, as well as applications (see also Rousseeuw and Leroy, 1987). However, LMS has an abnormally slow convergence rate and hence its asymptotic efficiency is zero. In contrast, LTS is asymptotically normal and can be computed much faster. The other high-breakdown regression method was Siegel's repeated median technique (Siegel, 1982), which has good properties in the simple regression case ($p = 2$) but is no longer affine equivariant in multiple regression ($p \geq 3$).

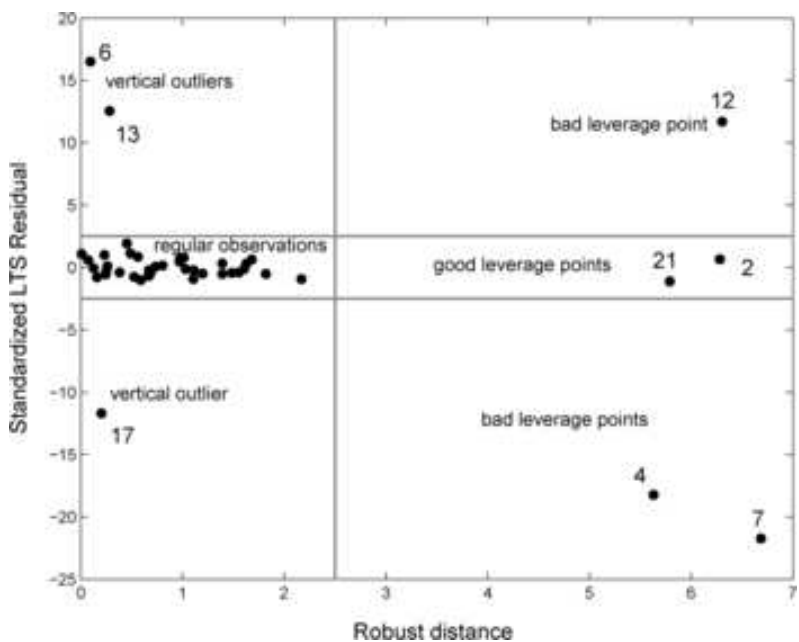

Fig. 5. *Regression outlier map of the data in Figure 4.*

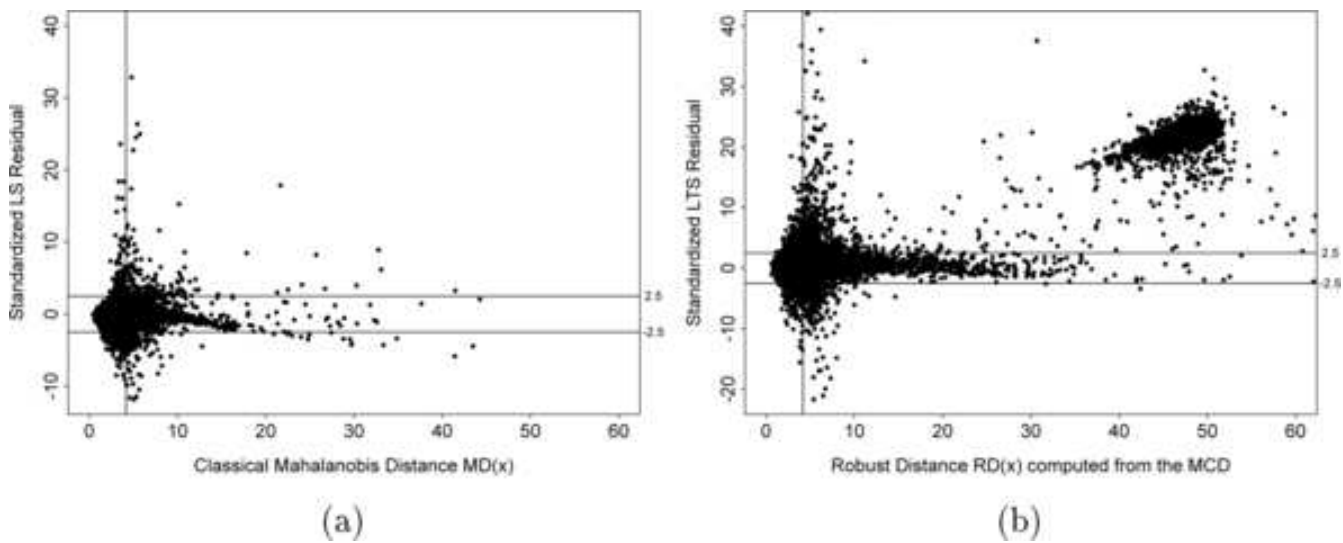

Fig. 6. *Digitized Palomar Sky Survey data: regression of MaperF on eight regressors. (a) Plot of LS residual versus Mahalanobis distance $MD(\mathbf{x}_i)$; (b) outlier map of LTS residual versus robust distance $RD(\mathbf{x}_i)$.*

As for multiple location and scatter, the efficiency of a high-breakdown regression estimator can be improved by computing one-step weighted least squares estimates (14)–(15) or by computing a one-step M-estimator as done in Rousseeuw (1984). In order to combine these advantages with those of the bounded influence approach, it was later proposed by Simpson, Ruppert and Carroll (1992), Coakley and Hettmansperger (1993) and Simpson and Yohai (1998) to compute a one-step GM-estimator starting from LTS.

Tests and variable selection for robust regression were developed by Markatou and He (1994), Markatou and Hettmansperger (1990), Ronchetti and Staudte (1994) and Ronchetti, Field and Blanchard (1997). For high-breakdown methods, variable selection by all subsets regression becomes infeasible. One way out is to apply the robust method to all variables, yielding weights, and then to apply the classical selection methods for weighted least squares. Alternatively, a robust $R^2$ measure (Croux and Dehon, 2003) or a robust penalized selection criterion (Müller and Welsh, 2006) can be used in a forward or backward selection strategy.

Another approach to improve the efficiency of the LTS is to replace its objective function by a smoother alternative. Similarly as in (8), S-estimators of regression (Rousseeuw and Yohai, 1984) are defined as the solution $(\hat{\boldsymbol{\theta}}, \hat{\sigma})$ that minimizes $\hat{\sigma}$ subject to the constraint

$$(16) \qquad \frac{1}{n} \sum_{i=1}^{n} \rho\left(\frac{y_i - \boldsymbol{\theta}' \mathbf{x}_i}{\sigma}\right) \leq b.$$

The constant $b$ usually equals $E_\Phi[\rho(Y)]$ to assure consistency at the model with normal error distribution, and as before $\rho$ is often taken to be Tukey's biweight $\rho$ function (9). Salibian-Barrera and Yohai (2006) recently constructed an efficient algorithm to compute regression S-estimators. Related classes of efficient high-breakdown estimators include MM-estimators (Yohai, 1987), $\tau$-estimators (Yohai and Zamar, 1988), a new type of R-estimators (Hössjer, 1994), generalized S-estimators (Croux, Rousseeuw and Hössjer, 1994), CM-estimators (Mendes and Tyler, 1996) and generalized $\tau$-estimators (Ferretti et al., 1999). Inference for these estimators is usually based on their asymptotic distribution at the central model. Alternatively, for MM-estimators Salibian-Barrera and Zamar (2002) developed a fast and robust bootstrap procedure that yields reliable nonparametric robust inference.

To extend the good properties of the univariate median to regression, Rousseeuw and Hubert (1999) introduced the notions of regression depth and deepest regression. The deepest regression estimator has been studied by Rousseeuw, Van Aelst and Hubert (1999b), Van Aelst and Rousseeuw (2000), Van Aelst et al. (2002) and Bai and He (2000).

Another important robustness measure, besides the breakdown value and the influence function, is the maxbias curve. The maxbias is the maximum possible bias of an estimator caused by a fixed fraction $\varepsilon$ of contamination. The maxbias curve plots the maxbias of an estimator as a function of the fraction $\varepsilon = m/n$ of contamination. Maxbias curves of robust regression estimators have been studied in Martin, Yohai and Zamar (1989), He and Simpson (1993), Croux, Rousseeuw and Hössjer (1994), Adrover and Zamar (2004) and Berrendero and Zamar (2001). Projection estimators for regression (Maronna and Yohai, 1993) combine a low maxbias with high breakdown value and bounded influence but they are difficult to compute.

Unbalanced binary regressors that contain, for example, 90% of zeroes and 10% of ones might be ignored by standard robust regression methods. Robust methods for regression models that include categorical or binary regressors have been developed by Hubert and Rousseeuw (1996) and Maronna and Yohai (2000). Robust estimators for orthogonal regression and error-in-variables models have been considered by Zamar (1989, 1992) and Maronna (2005).

## 4. MULTIVARIATE REGRESSION

The regression model can be extended to the case where we have more than one response variable. For $p$-variate predictors $\mathbf{x}_i = (x_{i1}, \ldots, x_{ip})'$ and $q$-variate responses $\mathbf{y}_i = (y_{i1}, \ldots, y_{iq})'$ the multivariate regression model is given by

$$(17) \qquad \mathbf{y}_i = \boldsymbol{\mathcal{B}}' \mathbf{x}_i + \boldsymbol{\alpha} + \boldsymbol{\varepsilon}_i,$$

where $\boldsymbol{\mathcal{B}}$ is the $p \times q$ slope matrix, $\boldsymbol{\alpha}$ is the $q$-dimensional intercept vector, and the errors $\boldsymbol{\varepsilon}_i = (\varepsilon_{i1}, \ldots, \varepsilon_{iq})'$ are i.i.d. with zero mean and with $\mathrm{Cov}(\boldsymbol{\varepsilon}) = \boldsymbol{\Sigma}_\varepsilon$ a positive definite matrix of size $q$. Note that for $q = 1$ we obtain the multiple regression model of the previous section. On the other hand, putting $p = 1$ and $x_i = 1$ yields the multivariate location and scatter model of Section 2. It is well known that the least squares solution can be written as

$$(18) \qquad \begin{aligned} \hat{\boldsymbol{\mathcal{B}}} &= \hat{\boldsymbol{\Sigma}}_{xx}^{-1} \hat{\boldsymbol{\Sigma}}_{xy}, \\ \hat{\boldsymbol{\alpha}} &= \hat{\boldsymbol{\mu}}_y - \hat{\boldsymbol{\mathcal{B}}}' \hat{\boldsymbol{\mu}}_x, \\ \hat{\boldsymbol{\Sigma}}_\varepsilon &= \hat{\boldsymbol{\Sigma}}_{yy} - \hat{\boldsymbol{\mathcal{B}}}' \hat{\boldsymbol{\Sigma}}_{xx} \hat{\boldsymbol{\mathcal{B}}}, \end{aligned}$$

where

$$\hat{\boldsymbol{\mu}} = \begin{pmatrix} \hat{\boldsymbol{\mu}}_x \\ \hat{\boldsymbol{\mu}}_y \end{pmatrix} \quad \text{and} \quad \hat{\boldsymbol{\Sigma}} = \begin{pmatrix} \hat{\boldsymbol{\Sigma}}_{xx} & \hat{\boldsymbol{\Sigma}}_{xy} \\ \hat{\boldsymbol{\Sigma}}_{yx} & \hat{\boldsymbol{\Sigma}}_{yy} \end{pmatrix}$$

are the empirical mean and covariance matrix of the joint $(\mathbf{x}, \mathbf{y})$ variables.

Vertical outliers and bad leverage points highly influence the least squares estimates in multivariate regression, and may make the results completely unreliable. Therefore, robust alternatives have been developed.

Rousseeuw et al. (2004) proposed to use the MCD estimates for the center $\boldsymbol{\mu}$ and scatter matrix $\boldsymbol{\Sigma}$ in (18). The resulting estimates are called MCD regression estimates. It has been shown that the MCD regression estimates are regression, $y$-affine and $x$-affine equivariant. With $\mathbf{X} = (\mathbf{x}_1, \dots, \mathbf{x}_n)'$, $\mathbf{Y} = (\mathbf{y}_1, \dots, \mathbf{y}_n)'$ and $\hat{\boldsymbol{\theta}} = (\hat{\boldsymbol{\mathcal{B}}}', \hat{\boldsymbol{\alpha}})'$ this means that

$$
\begin{aligned}
(19) \quad & \hat{\boldsymbol{\theta}}(\mathbf{X}, \mathbf{Y} + \mathbf{X}\mathbf{D} + \mathbf{1}_n \mathbf{w}') \\
& \quad = \hat{\boldsymbol{\theta}}(\mathbf{X}, \mathbf{Y}) + (\mathbf{D}', \mathbf{w})', \\
& \hat{\boldsymbol{\theta}}(\mathbf{X}, \mathbf{Y}\mathbf{C} + \mathbf{1}_n \mathbf{w}') \\
& \quad = \hat{\boldsymbol{\theta}}(\mathbf{X}, \mathbf{Y})\,\mathbf{C} + (\mathbf{O}'_{pq}, \mathbf{w})', \\
& \hat{\boldsymbol{\theta}}(\mathbf{X}\mathbf{A}' + \mathbf{1}_n \mathbf{v}', \mathbf{Y}) \\
& \quad = (\hat{\boldsymbol{\mathcal{B}}}'(\mathbf{X}, \mathbf{Y})\mathbf{A}^{-1}, \hat{\boldsymbol{\alpha}}(\mathbf{X}, \mathbf{Y}) \\
& \qquad - \hat{\boldsymbol{\mathcal{B}}}'(\mathbf{X}, \mathbf{Y})\mathbf{A}^{-1}\mathbf{v})',
\end{aligned}
$$

where $\mathbf{D}$ is any $p \times q$ matrix, $\mathbf{A}$ is any nonsingular $p \times p$ matrix, $\mathbf{C}$ is any nonsingular $q \times q$ matrix, $\mathbf{v}$ is any $p$-dimensional vector and $\mathbf{w}$ is any $q$-dimensional vector. Here $\mathbf{O}_{pq}$ is the $p \times q$ matrix consisting of zeroes.

MCD regression inherits the breakdown value of the MCD estimator, thus $\varepsilon_n^*(\hat{\boldsymbol{\theta}}) \approx (n - h)/n$. To obtain a better efficiency, the one-step weighted MCD estimates are used in (18) and followed by the regression weighting step described below. For any fit $\hat{\boldsymbol{\theta}}$ denote the corresponding $q$-dimensional residuals by $\mathbf{r}_i(\hat{\boldsymbol{\theta}}) = \mathbf{y}_i - \hat{\boldsymbol{\mathcal{B}}}'\mathbf{x}_i - \hat{\boldsymbol{\alpha}}$. Then the weighted regression estimates are given by

$$(20) \quad \hat{\boldsymbol{\theta}}_1 = \left( \sum_{i=1}^n w_i \mathbf{u}_i \mathbf{u}_i' \right)^{-1} \left( \sum_{i=1}^n w_i \mathbf{u}_i \mathbf{y}_i' \right),$$

$$(21) \quad \hat{\boldsymbol{\Sigma}}_\varepsilon^1 = d_1 \left( \sum_{i=1}^n w_i \right)^{-1} \left( \sum_{i=1}^n w_i \mathbf{r}_i(\hat{\boldsymbol{\theta}}_1) \mathbf{r}_i(\hat{\boldsymbol{\theta}}_1)' \right),$$

where $\mathbf{u}_i = (\mathbf{x}_i', 1)'$ and $d_1$ is a consistency factor. The weights $w_i$ are given by

$$w_i = \begin{cases} 1, & \text{if } d(\mathbf{r}_i(\hat{\boldsymbol{\theta}}_{\mathrm{MCD}})) \leq \sqrt{\chi^2_{q,0.975}}, \\ 0, & \text{otherwise}, \end{cases}$$

with $d(\mathbf{r}_i(\hat{\boldsymbol{\theta}}_{\mathrm{MCD}})) = \sqrt{\mathbf{r}_i(\hat{\boldsymbol{\theta}}_{\mathrm{MCD}})'(\hat{\boldsymbol{\Sigma}}_\varepsilon)^{-1}\mathbf{r}_i(\hat{\boldsymbol{\theta}}_{\mathrm{MCD}})}$ the robust distances of the residuals, corresponding to the initial MCD regression estimates $\hat{\boldsymbol{\theta}}_{\mathrm{MCD}}$ and $\hat{\boldsymbol{\Sigma}}_\varepsilon$. Note that these weighted regression estimates (20)–(21) have the same breakdown value as the initial MCD regression estimates.

To illustrate MCD regression we analyze a dataset from Shell's polymer laboratory, described in Rousseeuw et al. (2004). The dataset consists of $n = 217$ observations with $p = 4$ predictor variables and $q = 3$ response variables. The predictor variables describe the chemical characteristics of a piece of foam, whereas the response variables measure its physical properties such as tensile strength. The physical properties of the foam are determined by the chemical composition used in the production process. Multivariate regression is used to establish a relationship between the chemical inputs and the resulting physical properties of the foam. After an initial exploratory study of the variables, a robust multivariate MCD regression was used.

To detect leverage points and outliers the outlier map of Rousseeuw and van Zomeren (1990) has been extended to multivariate regression. In multivariate regression the robust distances of the residuals $\mathbf{r}_i(\hat{\boldsymbol{\theta}}_1)$ are plotted versus the robust distances of the $\mathbf{x}_i$. Figure 7 is the outlier map of the Shell foam data. Observations 215 and 110 lie far from both the horizontal cutoff line at $\sqrt{\chi^2_{3,0.975}} = 3.06$ and the vertical cutoff line at $\sqrt{\chi^2_{4,0.975}} = 3.34$. These two observations can thus be classified as bad leverage points. Several observations lie substantially above the horizontal cutoff but not to the right of the vertical cutoff, which means that they are vertical outliers (their residuals are outlying but their $x$-values are not).

Based on this list of special points the scientists who had made the measurements found out that a fraction of the observations in Figure 7 were made with a different production technique and hence belong to a different population with other characteristics. These include the observations 210, 212 and 215. We therefore remove these observations from the data, and retain only observations from the intended population.

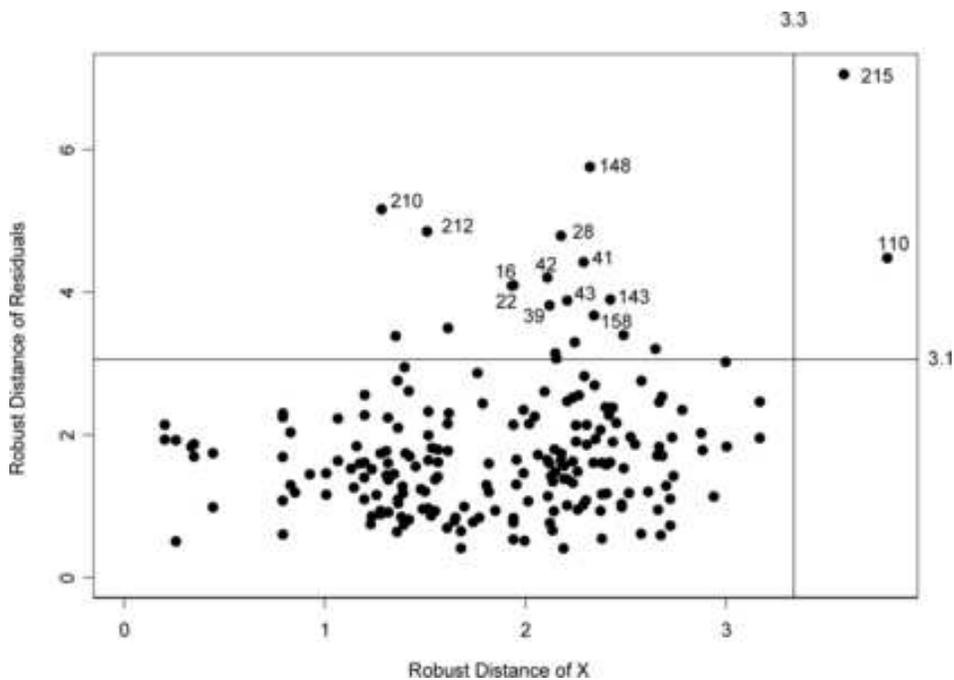

Fig. 7.  *Regression outlier map of the foam data.*

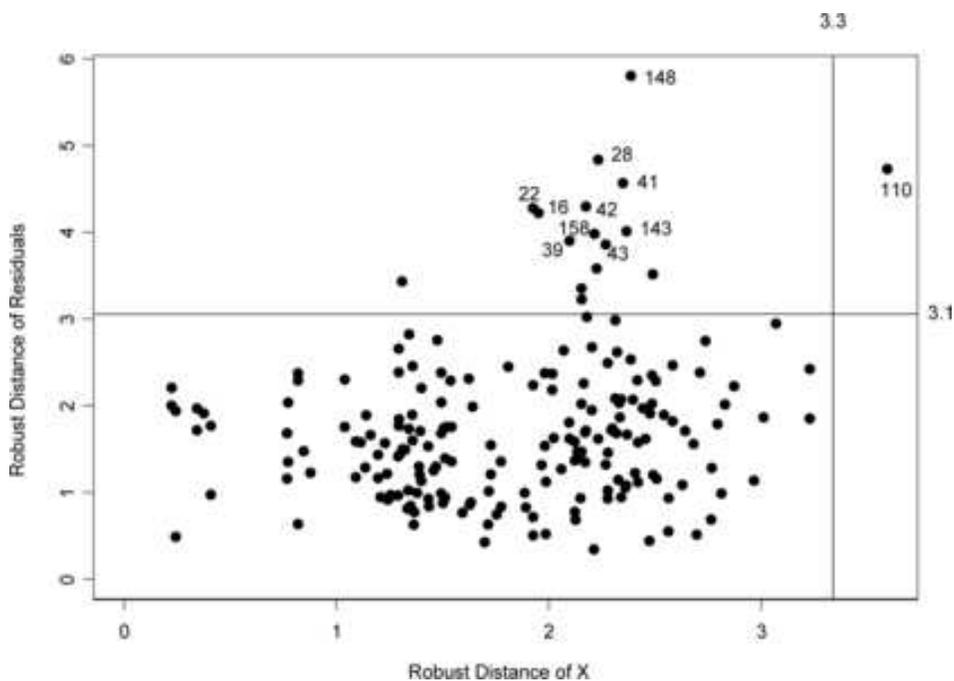

Fig. 8.  *Regression outlier map of the corrected foam data.*

Running the method again yields the outlier map in Figure 8. Observation 110 is still a bad leverage point, and also several of the vertical outliers remain. No chemical/physical mechanism was found to explain why these points are outliers, leaving open the possibility of some large measurement errors. But the detection of these outliers at least provides us with the option to choose whether or not to allow them to affect the final result.

Since MCD regression is mainly intended for regression data with random carriers, Agulló, Croux and Van Aelst (2006) developed an alternative robust multivariate regression method which can be seen as an extension of LTS to the multivariate setting. This multivariate least trimmed squares estimator (MLTS) can also be used in cases where the carriers are fixed. The MLTS looks for a subset of size $h$ such that the determinant of the covariance matrix of its residuals corresponding to its least squares fit is minimal. Similarly as for MCD regression, the MLTS has breakdown value $\varepsilon_n^*(\boldsymbol{\theta}_{\mathrm{MLTS}}) \approx (n-h)/n$ and the equivariance properties (19) are satisfied. The MLTS can be computed quickly with an algorithm similar to that in Appendix A.1. To improve the efficiency while keeping the breakdown value, a one-step weighted MLTS estimator can be computed using expressions (20)–(21). Alternatively, Van Aelst and Willems (2005) introduced multivariate regression S-estimators and extended the fast robust bootstrap methodology of Salibian-Barrera and Zamar (2002) to this setting while García Ben, Martínez and Yohai (2006) proposed $\tau$-estimators for multivariate regression.

## 5. CLASSIFICATION

The goal of classification, also known as discriminant analysis or supervised learning, is to obtain rules that describe the separation between known groups of observations. Moreover, it allows to classify new observations into one of the groups. We denote the number of groups by $l$ and assume that we can describe our experiment in each population $\pi_j$ by a $p$-dimensional random variable $X_j$ with density function $f_j$. We write $p_j$ for the membership probability, that is, the probability for an observation to come from $\pi_j$. The maximum likelihood rule then classifies an observation $\mathbf{x}$ into $\pi_k$ if $\ln(p_k f_k(\mathbf{x}))$ is the maximum of the set $\{\ln(p_j f_j(\mathbf{x})); j = 1, \ldots, l\}$. If we assume that the density $f_j$ for each group is Gaussian with mean $\boldsymbol{\mu}_j$ and covariance matrix $\boldsymbol{\Sigma}_j$, then it can be seen that the maximum likelihood rule is equivalent to maximizing the discriminant scores $d_j^Q(\mathbf{x})$ with

$$
\begin{aligned}
(22) \quad d_j^Q(\mathbf{x}) = {}& -\tfrac{1}{2}\ln|\boldsymbol{\Sigma}_j| \\
& -\tfrac{1}{2}(\mathbf{x}-\boldsymbol{\mu}_j)'\boldsymbol{\Sigma}_j^{-1}(\mathbf{x}-\boldsymbol{\mu}_j) \\
& + \ln(p_j).
\end{aligned}
$$

That is, $\mathbf{x}$ is allocated to $\pi_k$ if $d_k^Q(\mathbf{x}) > d_j^Q(\mathbf{x})$ for all $j = 1, \ldots, l$ (see, e.g., Johnson and Wichern, 1998).

In practice $\boldsymbol{\mu}_j$, $\boldsymbol{\Sigma}_j$ and $p_j$ have to be estimated. Classical quadratic discriminant analysis (CQDA) uses the group's mean and empirical covariance matrix to estimate $\boldsymbol{\mu}_j$ and $\boldsymbol{\Sigma}_j$. The membership probabilities are usually estimated by the relative frequencies of the observations in each group, hence $\hat{p}_j = n_j/n$ with $n_j$ the number of observations in group $j$.

A robust quadratic discriminant analysis (RQDA) is derived by using robust estimators of $\boldsymbol{\mu}_j$, $\boldsymbol{\Sigma}_j$ and $p_j$. In particular, we can apply the weighted MCD estimator of location and scatter in each group. As a byproduct of this robust procedure, outliers (within each group) can be distinguished from the regular observations. Finally, the membership probabilities can be robustly estimated as the relative frequency of *regular* observations in each group. For an outline of this approach, see Hubert and Van Driessen (2004).

When the groups are assumed to have a common covariance matrix $\Sigma$, the quadratic scores (22) can be simplified to

$$
(23) \quad d_j^L(\mathbf{x}) = \boldsymbol{\mu}_j'\boldsymbol{\Sigma}^{-1}\mathbf{x} - \tfrac{1}{2}\boldsymbol{\mu}_j'\boldsymbol{\Sigma}^{-1}\boldsymbol{\mu}_j + \ln(p_j)
$$

since the terms $-\tfrac{1}{2}\ln|\boldsymbol{\Sigma}|$ and $-\tfrac{1}{2}\mathbf{x}'\boldsymbol{\Sigma}^{-1}\mathbf{x}$ do not depend on $j$. The resulting scores (23) are linear in $\mathbf{x}$, hence the maximum likelihood rule belongs to the class of *linear discriminant analysis*. It is well known that if we have only two populations ($l = 2$) with a common covariance structure and if both groups have equal membership probabilities, this rule coincides with Fisher's linear discriminant rule. Robust linear discriminant analysis based on the MCD estimator or S-estimators has been studied in Hawkins and McLachlan (1997), He and Fung (2000), Croux and Dehon (2001) and Hubert and Van Driessen (2004). The latter paper computes $\hat{\boldsymbol{\mu}}_j$ and $\hat{\boldsymbol{\Sigma}}_j$ by weighted MCD and then defines the pooled covariance matrix $\hat{\boldsymbol{\Sigma}} = (\sum_{j=1}^{l} n_j \hat{\boldsymbol{\Sigma}}_j)/n$.

We consider a dataset that contains the spectra of three different cultivars of the same fruit (cantaloupe—Cucumis melo L. Cantaloupensis). The cultivars (named D, M and HA) have sizes 490, 106 and 500, and all spectra were measured in 256 wavelengths. The dataset thus contains 1096 observations and 256 variables. First, a robust principal component analysis (as described in the next section) was applied to reduce the dimension of the data space, and the first two components were retained. For a more detailed description and analysis of these data, see Hubert and Van Driessen (2004).

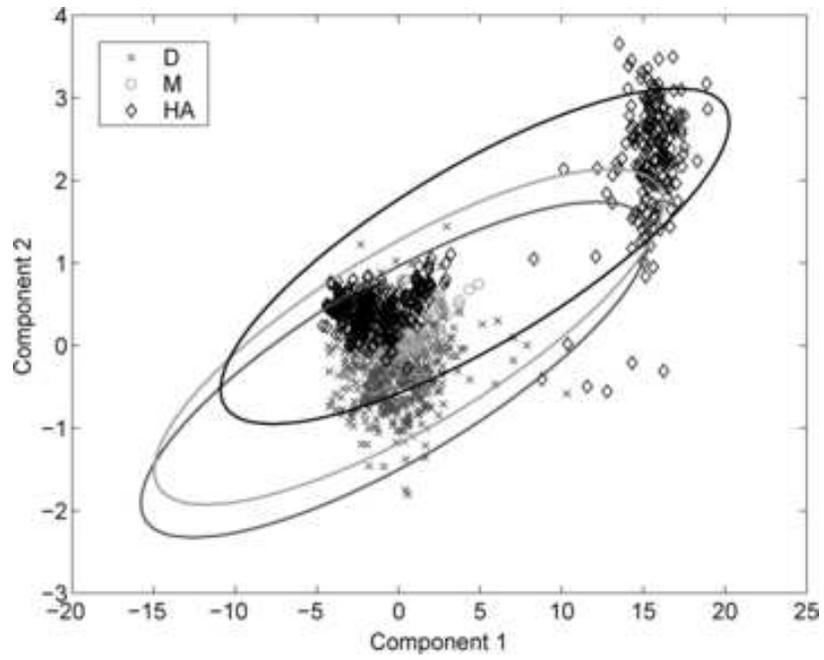

(a)

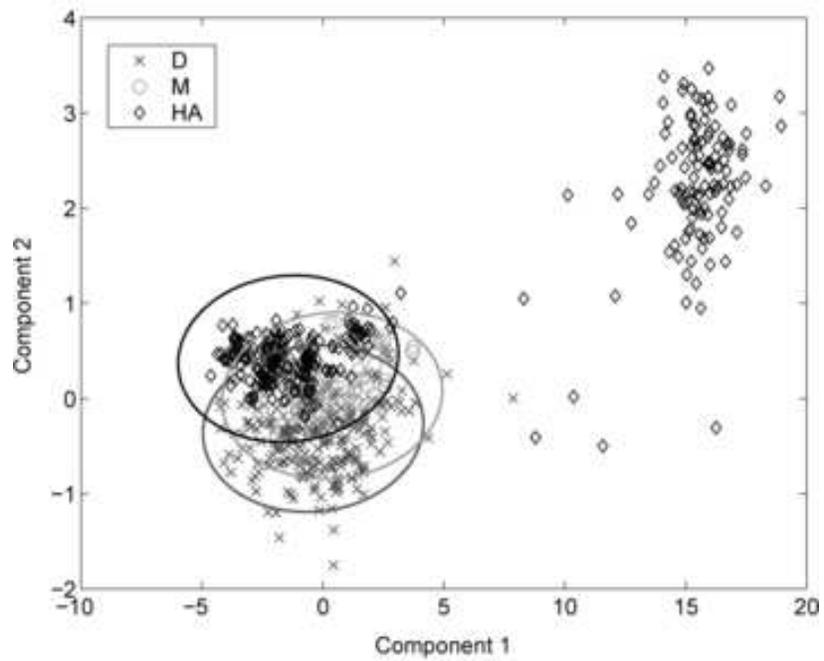

(b)

FIG. 9.   *(a) Classical tolerance ellipses for the fruit data with common covariance matrix; (b) robust tolerance ellipses.*

The data were divided randomly in a training set and a validation set, containing 60% and 40% of the observations. Figure 9 shows the training data. In this figure cultivar D is marked with crosses, cultivar M with circles and cultivar HA with diamonds. We see that cultivar HA has a cluster of outliers that are far away from the other observations. As it turns out, these outliers were caused by a change in the illumination system. To classify the data, we will use model (23) with a common covariance matrix $\boldsymbol{\Sigma}$. Figure 9(a) shows the classical tolerance ellipses for the groups, given by $(\mathbf{x} - \hat{\boldsymbol{\mu}}_j)' \hat{\boldsymbol{\Sigma}}^{-1} (\mathbf{x} - \hat{\boldsymbol{\mu}}_j) = \chi^2_{2,0.975}$. Note how strongly the classical covariance estimator of the common $\boldsymbol{\Sigma}$ is influenced by the outlying subgroup of cultivar HA. On the other hand, Figure 9(b) shows the same data with the corresponding robust tolerance ellipses.

The effect on the resulting classical linear discriminant rules is dramatic for cultivar M. It appears that all the observations are badly classified because they would have to belong to a region that lies completely outside the boundary of this figure! The robust discriminant analysis does a better job. The tolerance ellipses are not affected by the outliers and the resulting discriminant lines split up the different groups more accurately. The misclassification rates are 17% for cultivar D, 95% for cultivar M and 6% for cultivar HA. The misclassification rate of cultivar M remains very high. This is due to the intrinsic overlap between the three groups, and the fact that cultivar M has few data points compared to the others. (When we impose that all three groups are equally important by setting the membership probabilities equal to 1/3, we obtain a better classification of cultivar M with 46% of errors.)

This example thus clearly shows that outliers can have a huge effect on the classical discriminant rules, whereas the robust version fares better.

## 6. PRINCIPAL COMPONENT ANALYSIS

### 6.1 Classical PCA

Principal component analysis is a popular statistical method which tries to explain the covariance structure of data by means of a small number of components. These components are linear combinations of the original variables, and often allow for an interpretation and a better understanding of the different sources of variation. Because PCA is concerned with data reduction, it is widely used for the analysis of high-dimensional data which are frequently encountered in chemometrics, computer vision, engineering, genetics, and other domains. PCA is then often the first step of the data analysis, followed by discriminant analysis, cluster analysis, or other multivariate techniques (see, e.g., Hubert and Engelen, 2004). It is thus important to find those components that contain most of the information.

In the classical approach, the first component corresponds to the direction in which the projected observations have the largest variance. The second component is then orthogonal to the first and again maximizes the variance of the projected data points. Continuing in this way produces all the principal components, which correspond to the eigenvectors of the empirical covariance matrix. Unfortunately, both the classical variance (which is being maximized) and the classical covariance matrix (which is being decomposed) are very sensitive to anomalous observations. Consequently, the first components are often pulled toward outlying points, and may not capture the variation of the regular observations. Therefore, data reduction based on classical PCA (CPCA) becomes unreliable if outliers are present in the data.

To illustrate this, let us consider a small artificial dataset in $p = 4$ dimensions. The Hawkins–Bradu–Kass dataset (see, e.g., Rousseeuw and Leroy, 1987) consists of $n = 75$ observations in which two groups of outliers were created, labeled 1–10 and 11–14. The first two eigenvalues explain already 98% of the total variation, so we select $k = 2$. The CPCA scores plot is depicted in Figure 10(a). In this figure we can distinguish the two groups of outliers, but we see several other undesirable effects. We first observe that, although the scores have zero mean, the regular data points lie far from zero. This stems from the fact that the mean of the data points is a bad estimate of the true center of the data in the presence of outliers. It is clearly shifted toward the outlying group, and consequently the origin even falls outside the cloud of the regular data points. On the plot we have also superimposed the 97.5% tolerance ellipse. We see that the outliers 1–10 are within the tolerance ellipse, and thus do not stand out based on their Mahalanobis distance. The ellipse has stretched itself to accommodate these outliers.

### 6.2 Robust PCA

The goal of robust PCA methods is to obtain principal components that are not influenced much by outliers. A first group of methods is obtained by replacing the classical covariance matrix by a robust covariance estimator. Maronna (1976) and Campbell (1980) proposed using affine equivariant M-estimators of scatter for this purpose, but these cannot resist many outliers. Croux and Haesbroeck (2000) used positive-breakdown estimators of scatter such as the MCD and S-estimators. Recently, Salibian-Barrera, Van Aelst and Willems (2006) proposed using S- or MM-estimators of scatter and developed a fast robust bootstrap procedure for inference and to assess the stability of the PCA solution. Let us reconsider the Hawkins–Bradu–Kass data in $p = 4$ dimensions. Robust PCA using the weighted MCD estimator yields the score plot in Figure 10(b). We now see that the center is correctly estimated in the middle of the regular observations. The 97.5% tolerance ellipse nicely encloses these points and excludes all 14 outliers.

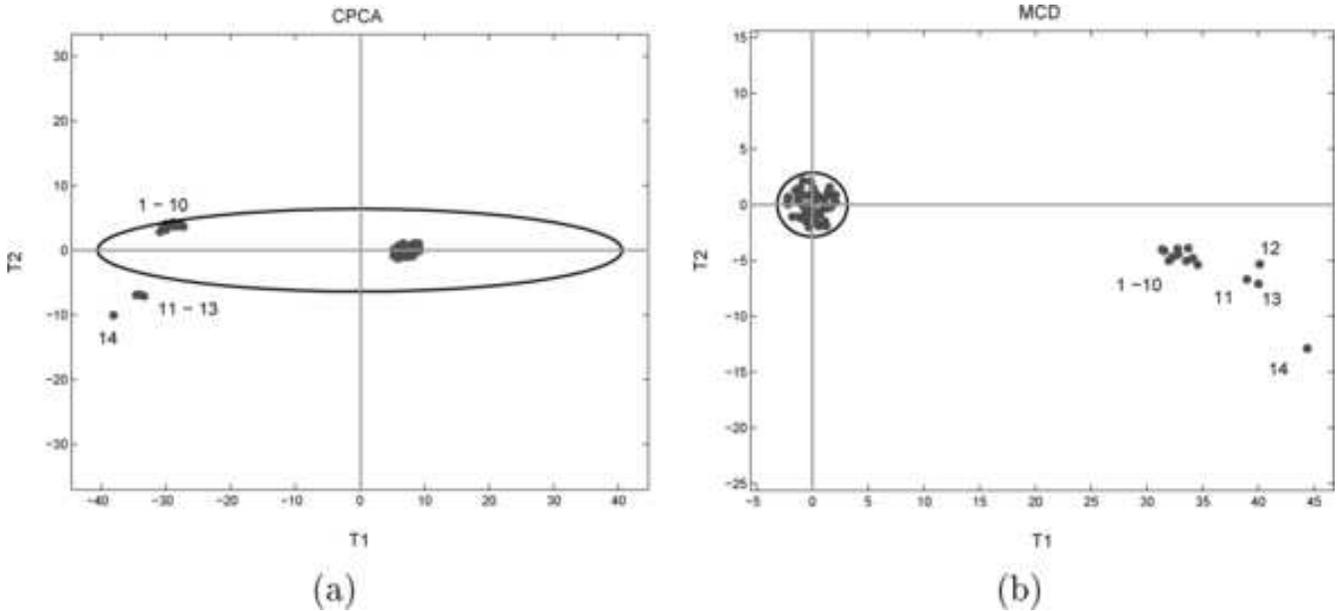

Fig. 10.  *Score plot and 97.5% tolerance ellipse of the Hawkins–Bradu–Kass data obtained with (a) CPCA; (b) MCD.*

Unfortunately, the use of these affine equivariant covariance estimators is limited to small to moderate dimensions. To see why, consider, for example, the MCD estimator. If $p$ denotes the number of variables in our dataset, the MCD estimator can only be computed if $p < h$; otherwise the covariance matrix of any $h$-subset has zero determinant. Since $h < n$, $p$ can never be larger than $n$. A second problem is the computation of these robust estimators in high dimensions. Today's fastest algorithms (Woodruff and Rocke, 1994; Rousseeuw and Van Driessen, 1999) can handle up to about 100 dimensions, whereas there are fields like chemometrics, which need to analyze data with dimensions in the thousands.

A second approach to robust PCA uses *projection pursuit* (PP) techniques. These methods maximize a robust measure of spread to obtain consecutive directions on which the data points are projected. In Hubert, Rousseeuw and Verboven (2002) a PP algorithm is presented, based on the ideas of Li and Chen (1985) and Croux and Ruiz-Gazen (1996, 2005). It has been successfully applied in several studies, for example, to detect outliers in large microarray data (Model et al., 2002). Asymptotic results about this approach are presented in Cui, He and Ng (2003).

Hubert, Rousseeuw and Vanden Branden (2005) proposed a robust PCA method, called ROBPCA, which combines ideas of both projection pursuit and robust covariance estimation. The PP part is used for the initial dimension reduction. Some ideas based on the MCD estimator are then applied to this lower-dimensional data space. Simulations in Hubert, Rousseeuw and Vanden Branden (2005) have shown that this combined approach yields more accurate estimates than the raw PP algorithm. An outline of the ROBPCA algorithm is given in Appendix A.3.

The ROBPCA method applied to a dataset $\mathbf{X}_{n,p}$ yields robust principal components which can be collected in a loading matrix $\mathbf{P}_{p,k}$ with orthogonal columns, and a robust center $\hat{\boldsymbol{\mu}}_x$. From here on the subscripts to a matrix serve to recall its size; fore example, $\mathbf{X}_{n,p}$ is an $n \times p$ matrix and $\mathbf{P}_{p,k}$ is $p \times k$. (Note that it is possible to robustly scale the variables first by dividing them by a robust scale estimate; see, e.g., Rousseeuw and Croux, [1993].) The robust scores are the $k \times 1$ column vectors

$$\mathbf{t}_i = (\mathbf{P}_{p,k})'(\mathbf{x}_i - \hat{\boldsymbol{\mu}}_x).$$

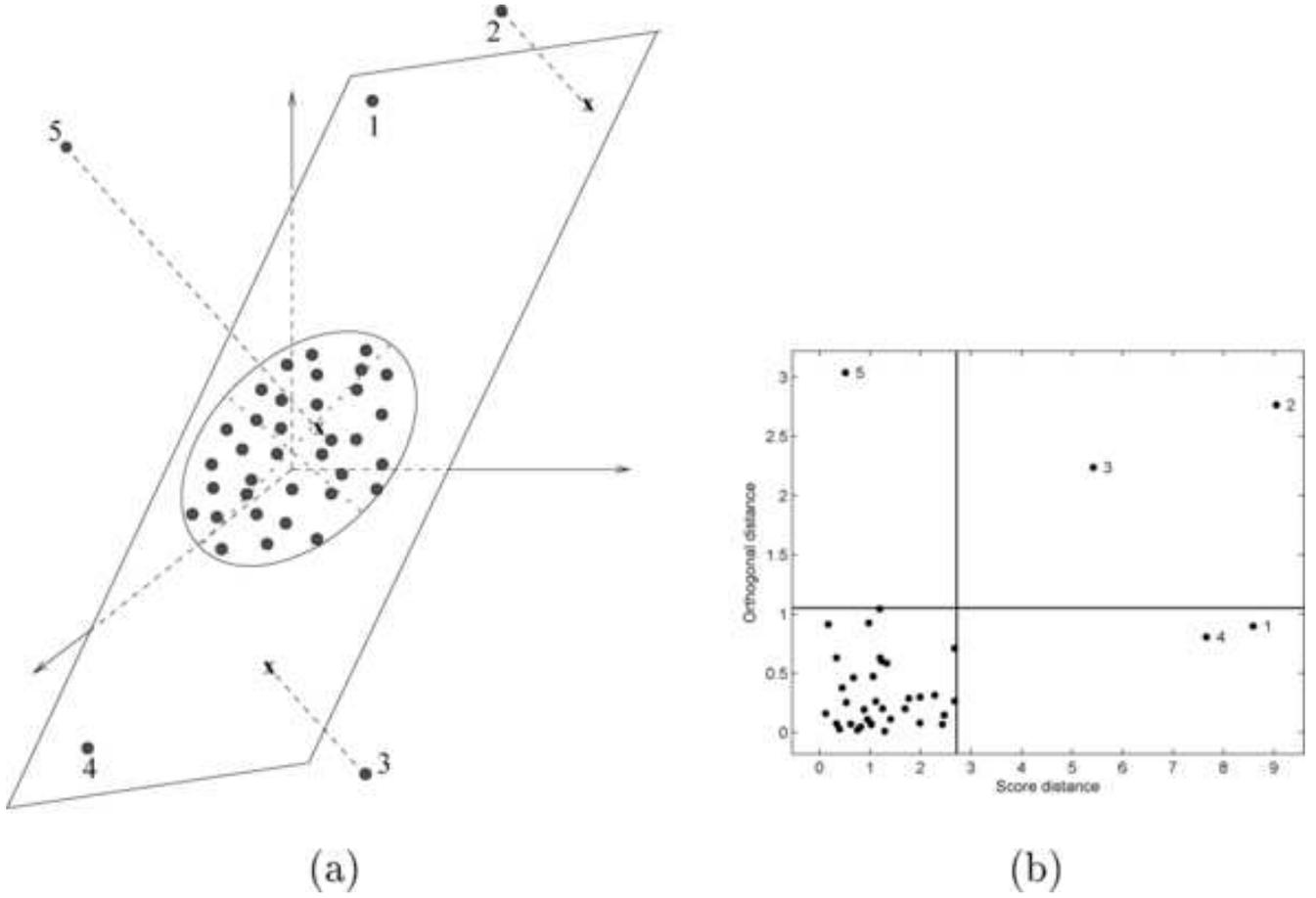

Fig. 11. (a) *Different types of outliers when a three-dimensional dataset is projected on a robust two-dimensional PCA-subspace; (b) the corresponding PCA outlier map.*

The *orthogonal distance* measures the distance between an observation and its projection in the $k$-dimensional PCA subspace:

$$(24) \qquad \mathrm{OD}_i = \|\mathbf{x}_i - \hat{\boldsymbol{\mu}}_x - \mathbf{P}_{p,k}\mathbf{t}_i\|.$$

Let $\mathbf{L}$ denote the diagonal matrix which contains the eigenvalues $l_j$ of the MCD scatter matrix, sorted from largest to smallest. The *score distance* of $\mathbf{x}_i$ with respect to $\hat{\boldsymbol{\mu}}_x, \mathbf{P}$ and $\mathbf{L}$ is then defined as

$$\mathrm{SD}_i = \sqrt{(\mathbf{x}_i - \hat{\boldsymbol{\mu}}_x)'\mathbf{P}_{p,k}\mathbf{L}_{k,k}^{-1}(\mathbf{P}_{p,k})'(\mathbf{x}_i - \hat{\boldsymbol{\mu}}_x)}$$

$$= \sqrt{\sum_{j=1}^{k} \frac{t_{ij}^2}{l_j}}.$$

All the above mentioned methods are translation and orthogonal equivariant, that is, (2)–(3) hold for any vector $\mathbf{v}$ and any $p \times p$ matrix $\mathbf{A}$ with $\mathbf{AA}' = \mathbf{I}$. To be precise, let $\hat{\boldsymbol{\mu}}_x$ and $\mathbf{P}$ denote the robust center and loading matrix of the original observations $\mathbf{x}_i$. Then the robust center and loadings of the transformed data $\mathbf{Ax}_i + \mathbf{v}$ are equal to $\mathbf{A}\hat{\boldsymbol{\mu}}_x + \mathbf{v}$ and $\mathbf{AP}$. The scores (and distances) remain the same after this transformation, since

$$\mathbf{t}_i(\mathbf{Ax}_i + \mathbf{v}) = \mathbf{P}'\mathbf{A}'(\mathbf{Ax}_i + \mathbf{v} - (\mathbf{A}\hat{\boldsymbol{\mu}}_x + \mathbf{v}))$$

$$= \mathbf{P}'(\mathbf{x}_i - \hat{\boldsymbol{\mu}}_x) = \mathbf{t}_i(\mathbf{x}_i).$$

We also mention the robust LTS-subspace estimator and its generalizations, introduced and discussed in Rousseeuw and Leroy ([1987]) and Maronna ([2005]). The idea behind these approaches consists in minimizing a robust scale of the orthogonal distances, similar to the LTS estimator and S-estimators in regression. For functional data, a fast PCA method is introduced in Locantore et al. (1999).

## 6.3 Outlier Map

The result of the PCA analysis can be represented by means of the outlier map given in Hubert, Rousseeuw and Vanden Branden (2005). As in regression, this figure highlights the outliers and classifies them into several types. In general, an outlier is an observation which does not obey the pattern of the majority of the data. In the context of PCA, this means that an outlier either lies far from the subspace spanned by the $k$ eigenvectors, and/or that the projected observation lies far from the bulk of the data within this subspace. This can be expressed by means of the orthogonal and the score distances. These two distances define four types of observations, as illustrated in Figure 11(a). *Regular observations* have a small orthogonal and a small score distance. *Bad leverage points*, such as observations 2 and 3, have a large orthogonal distance and a large score distance. They typically have a large influence on classical PCA, as the eigenvectors will be tilted toward them. When points have a large score distance but a small orthogonal distance, we call them *good leverage points*. Observations 1 and 4 in Figure 7(a) can be classified into this category. Finally, *orthogonal outliers* have a large orthogonal distance, but a small score distance, as, for example, case 5. They cannot be distinguished from the regular observations once they are projected onto the PCA subspace, but they lie far from this subspace.

The outlier map in Figure 11(b) displays the $OD_i$ versus the $SD_i$. In this plot, lines are drawn to distinguish the observations with a small and a large OD, and with a small and a large SD. For the latter distances, the cutoff value $c = \sqrt{\chi^2_{k,0.975}}$ is used. For the orthogonal distances, the approach of Box (1954) is followed. The squared orthogonal distances can be approximated by a scaled $\chi^2$ distribution which in its turn can be approximated by a normal distribution using the Wilson–Hilferty transformation. The mean and variance of this normal distribution are then estimated by applying the univariate MCD to the $OD_i^{2/3}$.

## 6.4 Example

We illustrate the PCA outlier map on a dataset consisting of spectra of 180 archaeological glass pieces over $p = 750$ wavelengths (Lemberge et al., 2000). The measurements were performed using a Jeol JSM 6300 scanning electron microscope equipped with an energy-dispersive Si(Li) X-ray detection system. Three principal components were retained for CPCA and ROBPCA, yielding the outlier maps in Figure 12. In Figure 12(a) we see that CPCA does not find big outliers. On the other hand the ROBPCA plot in Figure 12(b) clearly distinguishes two major groups in the data, as well as a smaller group of bad leverage points, a few orthogonal outliers, and the isolated case 180 in between the two major groups. A high-breakdown method such as ROBPCA detects the smaller group with cases 143–179 as a set of outliers. Later, it turned out that the window of the detector system had been cleaned before the last 38 spectra were measured. As a result less X-ray radiation was absorbed, resulting in higher X-ray intensities. The other bad leverage points (57–63) and (74–76) are samples with a large concentration of calcic. The orthogonal outliers (22, 23 and 30) are borderline cases, although it turned out that they have larger measurements at the channels 215–245. This might indicate a larger concentration of phosphorus.

## 7. PRINCIPAL COMPONENT REGRESSION

Principal component regression is typically used for linear regression models (10) or (17) where the number of independent variables $p$ is very large or where the regressors are highly correlated (this is known as multicollinearity). An important application of PCR is multivariate calibration in chemometrics, which predicts constituent concentrations of a material based on its spectrum. This spectrum can be obtained via several techniques such as fluorescence spectrometry, near-infrared spectrometry (NIR), nuclear magnetic resonance (NMR), ultraviolet spectrometry (UV), energy dispersive X-ray fluorescence spectrometry (ED-XRF), and so on. Since a spectrum typically ranges over a large number of wavelengths, it is a high-dimensional vector with hundreds of components. The number of concentrations, on the other hand, is usually limited to at most, say, five. In the univariate approach, only one concentration at a time is modeled and analyzed. The more general problem assumes that the number of response variables $q$ is larger than 1, which means that several concentrations are to be estimated together. This model has the advantage that the covariance structure between the concentrations is also taken into account, which is appropriate when the concentrations are known to be strongly correlated with each other.

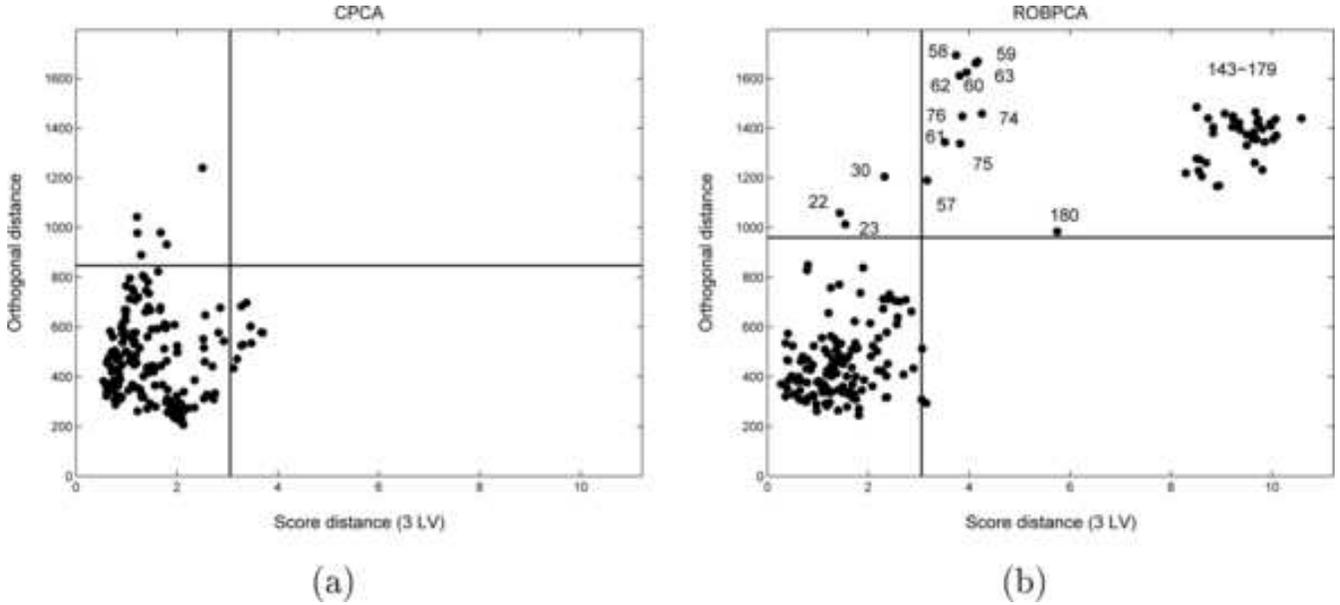

Fig. 12. *PCA outlier map of the glass dataset based on three principal components, computed with (*a*) CPCA; (*b*) ROBPCA.*

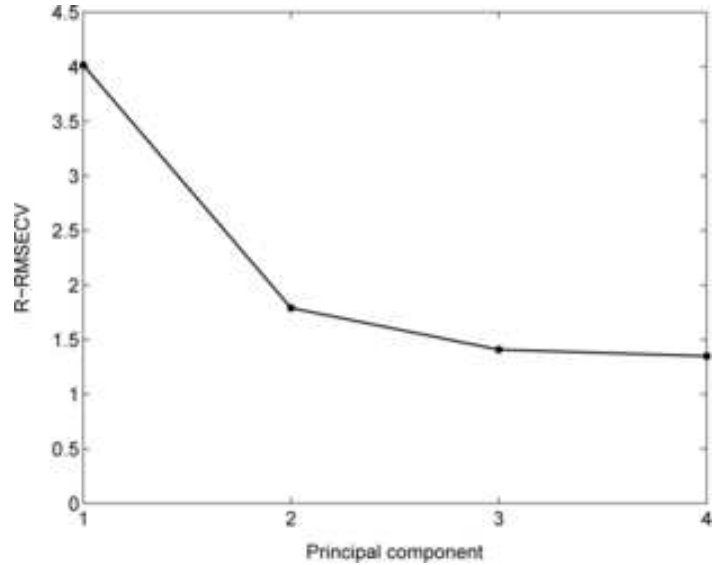

Fig. 13. *Robust R-RMSECV$_k$ curve for the Biscuit Dough dataset.*

Classical PCR (CPCR) starts by replacing the large number of explanatory variables $X_j$ by a small number of loading vectors, which correspond to the first (classical) principal components of $\mathbf{X}_{n,p}$. Then the response variables $Y_j$ are regressed on these components using least squares regression. It is thus a two-step procedure, which starts by computing scores $\mathbf{t}_i$ for every data point. Then the $\mathbf{y}_i$ are regressed on the $\mathbf{t}_i$.

The robust PCR method proposed by Hubert and Verboven (2003) combines robust PCA for high-dimensional $x$-data with a robust multivariate regression technique such as MCD regression described in Section 4. The robust scores $\mathbf{t}_i$ obtained with ROBPCA thus serve as the explanatory variables in the regression model (10) or (17).

The RPCR method inherits the $y$-affine equivariance [the second equation in (19)] from the MCD re-gression method. RPCR is also $x$-translation equivariant and $x$-orthogonally equivariant, that is, the estimates satisfy the third equation in (19) for any orthogonal matrix $\mathbf{A}$. These properties follow in a straightforward way from the orthogonal equivariance of the ROBPCA method. Robust PCR methods which are based on nonequivariant PCA estimators, such as those proposed in Pell (2000), are not $x$-equivariant.

An important issue in PCR is selecting the number of principal components, for which several methods have been proposed. A popular approach minimizes the root mean squared error of cross-validation criterion RMSECV$_k$ which, for one response variable ($q = 1$), equals

$$(25) \qquad \text{RMSECV}_k = \sqrt{\frac{1}{n} \sum_{i=1}^{n} (y_i - \hat{y}_{-i,k})^2}$$

with $\hat{y}_{-i,k}$ the predicted value for observation $i$, where $i$ was left out of the dataset when performing the PCR method with $k$ principal components. The goal of the RMSECV$_k$ statistic is twofold. It yields an estimate of the root mean squared prediction error $E(y - \hat{y})^2$ when $k$ components are used in the model, whereas the curve of RMSECV$_k$ for $k = 1, \dots, k_{\max}$ is a popular graphical tool to choose the optimal number of components.

This RMSECV$_k$ statistic is, however, not suited at contaminated datasets because it also includes the prediction error of the outliers in (25). Therefore Hubert and Verboven (2003) proposed a robust RMSECV measure. These R-RMSECV$_k$ values were rather time consuming, because for every choice of $k$ they required the whole RPCR procedure to be performed $n$ times. Faster algorithms for cross-validation have recently been developed (Engelen and Hubert, 2005). They avoid the complete recomputation of resampling methods such as the MCD when one observation is removed from the dataset.

To illustrate RPCR we analyze the Biscuit Dough dataset of Osborne et al. (1984), preprocessed as in Hubert, Rousseeuw and Verboven (2002). This dataset consists of 40 NIR spectra of biscuit dough with measurements every 2 nanometers, from 1200 nm up to 2400 nm. The responses are the percentages of four constituents in the biscuit dough: $y_1 =$ fat, $y_2 =$ flour, $y_3 =$ sucrose and $y_4 =$ water. Because there is a significant correlation among the responses, a multivariate regression is performed. The robust

R-RMSECV$_k$ curve is plotted in Figure 13 and suggests to select $k = 2$ components.

Differences between CPCR and RPCR show up in the loading vectors and in the calibration vectors. Figure 14 shows the second loading vector and the second calibration vector for $y_3$ (sucrose). For instance, CPCR and RPCR give quite different results between wavelengths 1390 and 1440 (the so-called C–H bend).

Next, we can construct outlier maps as in Sections 4 and 6.3. ROBPCA yields the PCA outlier map displayed in Figure 15(a). We see that there are no leverage points but there are some orthogonal outliers, the largest being 23, 7 and 20. The result of the regression step is shown in Figure 15(b). It plots the robust distances of the residuals (or the standardized residuals if $q = 1$) versus the score distances. RPCR shows that observation 21 has an extremely high residual distance. Other vertical outliers are 23, 7, 20 and 24, whereas there are a few borderline cases.

## 8. PARTIAL LEAST SQUARES REGRESSION

Partial least squares regression (PLSR) is similar to PCR. Its goal is to estimate regression coefficients in a linear model with a large number of $x$-variables which are highly correlated. In the first step of PCR, the scores were obtained by extracting the main information present in the $x$-variables by performing a principal component analysis on them, without using any information about the $y$-variables. In contrast, the PLSR scores are computed by maximizing a covariance criterion between the $x$- and $y$-variables. Hence, this technique uses the responses already from the start.

More precisely, let $\tilde{\mathbf{X}}_{n,p}$ and $\tilde{\mathbf{Y}}_{n,q}$ denote the mean-centered data matrices, with $\tilde{\mathbf{x}}_i = \mathbf{x}_i - \bar{\mathbf{x}}$ and $\tilde{\mathbf{y}}_i = \mathbf{y}_i - \bar{\mathbf{y}}$. The normalized PLS weight vectors $\mathbf{r}_a$ and $\mathbf{q}_a$ (with $\|\mathbf{r}_a\| = \|\mathbf{q}_a\| = 1$) are then defined as the vectors that maximize

$$(26) \quad \text{cov}(\tilde{\mathbf{Y}}\mathbf{q}_a, \tilde{\mathbf{X}}\mathbf{r}_a) = \mathbf{q}_a' \frac{\tilde{\mathbf{Y}}'\tilde{\mathbf{X}}}{n-1} \mathbf{r}_a = \mathbf{q}_a' \hat{\mathbf{\Sigma}}_{yx} \mathbf{r}_a$$

for each $a = 1, \dots, k$, where $\hat{\mathbf{\Sigma}}_{yx}' = \hat{\mathbf{\Sigma}}_{xy} = \frac{\tilde{\mathbf{X}}'\tilde{\mathbf{Y}}}{n-1}$ is the empirical cross-covariance matrix between the $x$- and the $y$-variables. The elements of the scores $\tilde{\mathbf{t}}_i$ are then defined as linear combinations of the mean-centered data: $\tilde{t}_{ia} = \tilde{\mathbf{x}}_i' \mathbf{r}_a$, or equivalently $\tilde{\mathbf{T}}_{n,k} = \tilde{\mathbf{X}}_{n,p} \mathbf{R}_{p,k}$ with $\mathbf{R}_{p,k} = (\mathbf{r}_1, \dots, \mathbf{r}_k)$.

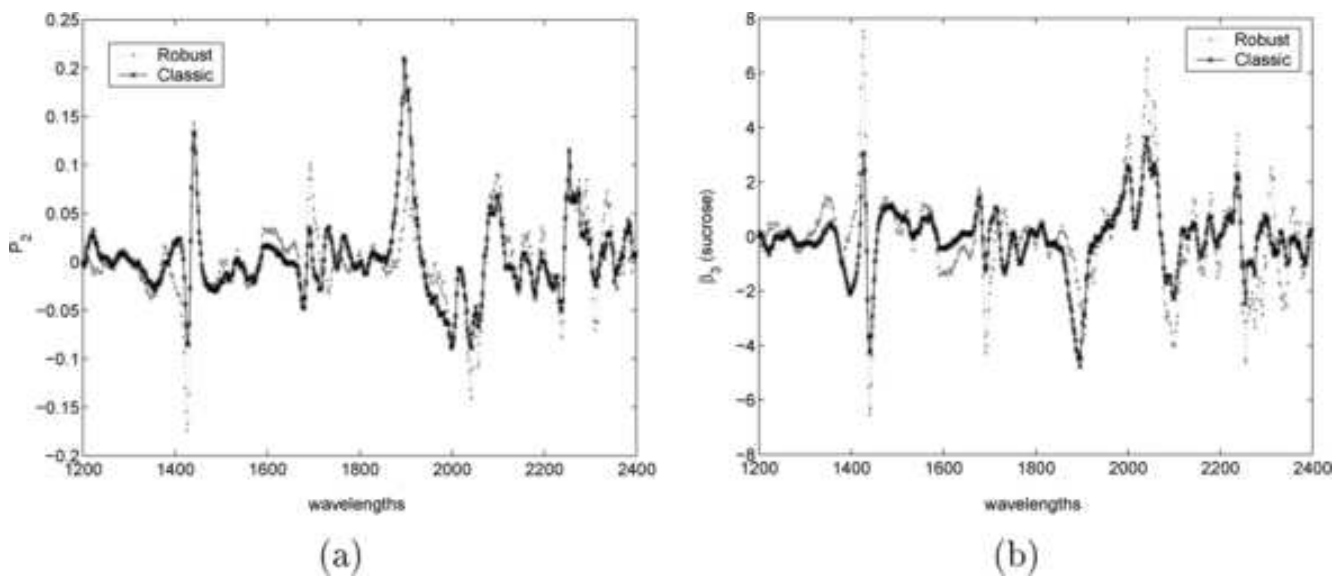

Fig. 14. *Second loading vector and calibration vector of sucrose for the Biscuit Dough dataset, computed with* (a) *CPCR;* (b) *RPCR.*

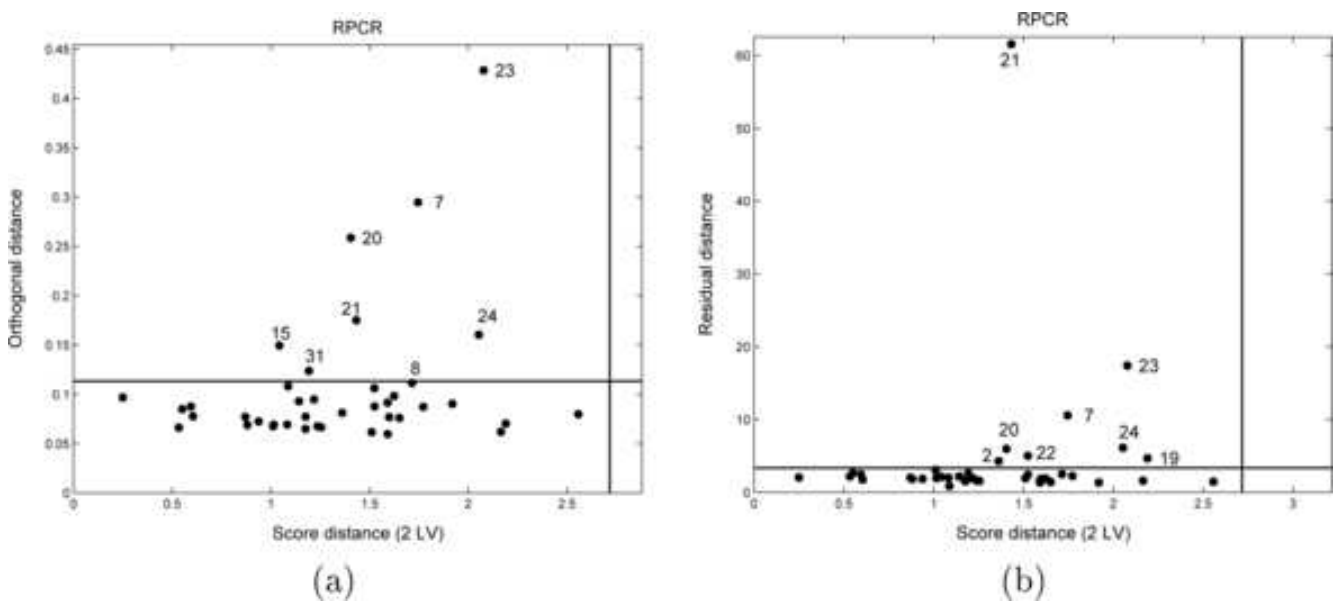

Fig. 15. (a) *PCA outlier map when applying RPCR to the Biscuit Dough dataset;* (b) *corresponding regression outlier map.*

The computation of the PLS weight vectors can be performed using the SIMPLS algorithm (de Jong, 1993), which is described in Appendix A.4.

Hubert and Vanden Branden (2003) developed the robust method RSIMPLS. It starts by applying ROBPCA on the $x$- and $y$-variables in order to replace $\hat{\boldsymbol{\Sigma}}_{xy}$ and $\hat{\boldsymbol{\Sigma}}_x$ by robust estimates, and then proceeds analogously to the SIMPLS algorithm. Similarly to RPCR, a robust regression method (ROBPCA regression) is performed in the second stage. Van-

den Branden and Hubert (2004) proved that for low-dimensional data the RSIMPLS approach yields bounded influence functions for the weight vectors $\mathbf{r}_a$ and $\mathbf{q}_a$ and for the regression estimates. Also the breakdown value is inherited from the MCD estimator.

The robustness of RSIMPLS is illustrated on the octane dataset (Esbensen, Schönkopf and Midtgaard, 1994), consisting of NIR absorbance spectra over $p = 226$ wavelengths ranging from 1102 nm to 1552 nm with measurements every two nanometers. For each of the $n = 39$ production gasoline samples the octane number $y$ was measured, so $q = 1$. It is known that the octane dataset contains six outliers (25, 26, 36–39) to which alcohol was added. From the RM-SECV values (Engelen et al., 2004) it follows that $k = 2$ components should be retained.

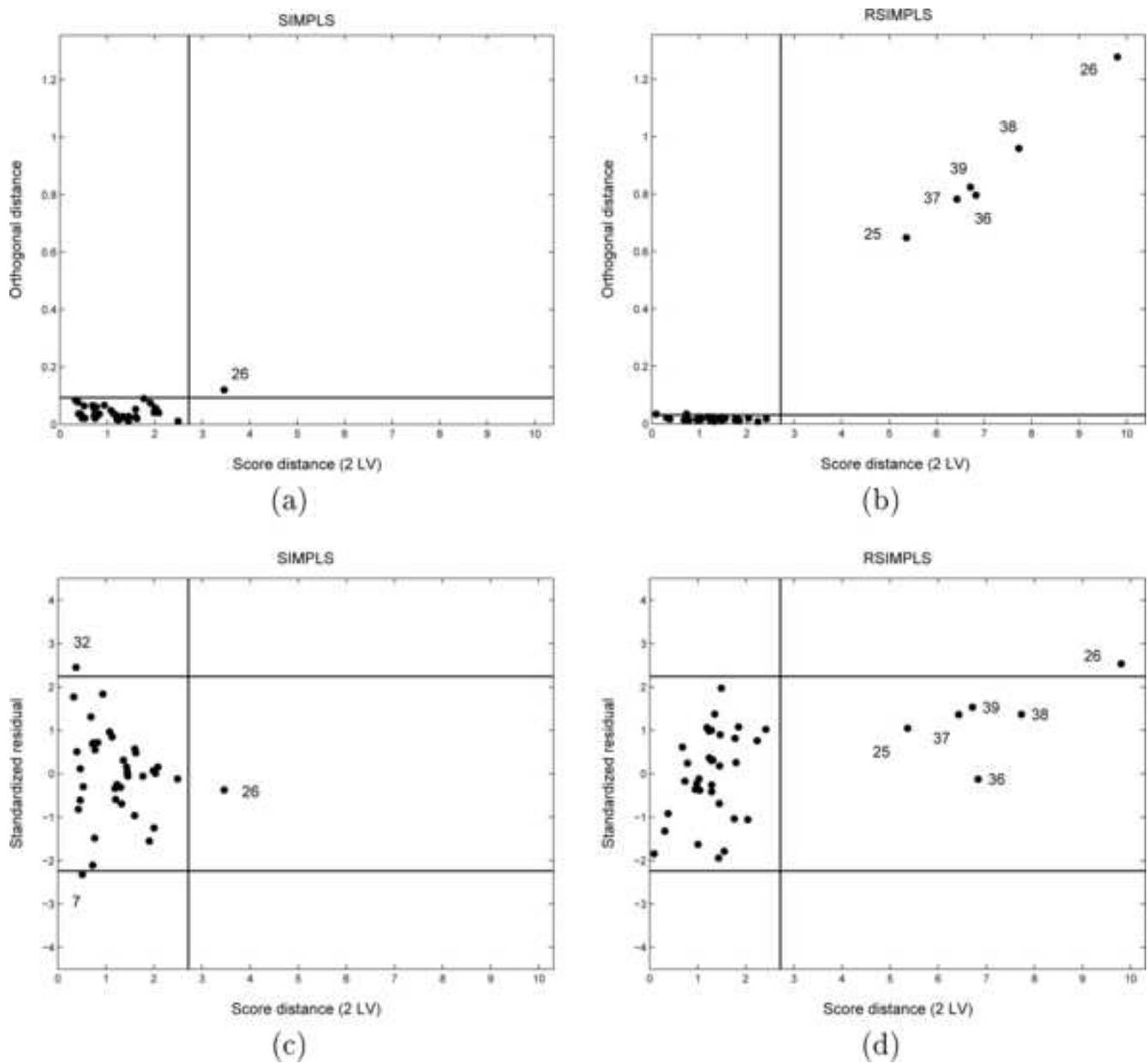

Fig. 16.   (a) *Score outlier map of the octane dataset using the SIMPLS results; (b) based on RSIMPLS; (c) regression outlier map based on SIMPLS; (d) based on RSIMPLS.*

The SIMPLS outlier map is Figure 16(a). We see that the classical analysis only detects the outlying spectrum 26, which does not even stick out much above the border line. The robust score outlier map is displayed in Figure 16(b). Here we immediately spot the six samples with added alcohol. The robust regression outlier map in Figure 16(d) shows that the outliers are good leverage points, whereas SIMPLS again only reveals spectrum 26.

Note that canonical correlation analysis tries to maximize the *correlation* between linear combinations of the $x$- and the $y$-variables, instead of the covariance in (26). Robust methods for canonical correlation are presented in Croux and Dehon (2002).

## 9. SOME OTHER MULTIVARIATE FRAMEWORKS

Apart from the frameworks covered in the previous sections, there is also work in other multivariate settings. These methods cannot be described in detail here due to lack of space, but here are some pointers to the literature. In the framework of multivariate location and scatter, an MCD-based alternative to the Hotelling test was provided by Willems et al. (2002) and a technique based on robust distances was applied to the control of electrical power systems in Mili et al. (1996). High-breakdown regression techniques were extended to computer vision settings (e.g., Meer et al., 1991; Stewart, 1995). For generalized linear models, robust approaches have been proposed by Cantoni and Ronchetti (2001), Künsch, Stefanski and Carroll (1989), Markatou, Basu and Lindsay (1998), Müller and Neykov (2003) and Rousseeuw and Christmann (2003). A high-breakdown method for mixed linear models has been proposed by Copt and Victoria-Feser (2006). Robust nonlinear regression methods have been studied by Stromberg (1993), Stromberg and Ruppert (1992) and Mizera (2002), who considered a depth-based approach. Boente, Pires and Rodrigues (2002) introduced robust estimators for common principal components. Robust methods were proposed for factor analysis (Pison et al., 2003) and independent component analysis (Brys, Hubert and Rousseeuw, 2005). Croux et al. (2003) fitted general multiplicative models such as FANOVA. Robust clustering methods have been investigated by Kaufman and Rousseeuw (1990), Cuesta-Albertos, Gordaliza and Matrán (1997) and Hardin and Rocke (2004). Robustness in time series analysis and econometrics has been studied by Martin and Yohai (1986), Bustos and Yohai (1986), Muler and Yohai (2002), Franses, Kloek and Lucas (1999), van Dijk, Franses and Lucas (1999a, 1999b) and Lucas and Franses (1998). Of course, this short list is far from complete.

## 10. AVAILABILITY

Stand-alone programs carrying out FAST-MCD and FAST-LTS can be downloaded from the website **http://www.agoras.ua.ac.be**, as well as Matlab versions. The FAST-MCD algorithm is available in the package S-PLUS (as the built-in function *cov.mcd*), in R (as part of the packages *rrcov*, *robust* and *robustbase*), and in SAS/IML Version 7. It is also included in SAS Version 9 (in *PROC ROBUSTREG*). These packages all provide the one-step weighted MCD estimates. The LTS is available in S-PLUS as the built-in function *ltsreg*, which uses a slower algorithm and has a low default breakdown value. The FAST-LTS algorithm is available in R (as part of *rrcov* and *robustbase*) and in SAS/IML Version 7. In SAS Version 9 it is incorporated in *PROC ROBUSTREG*.

Matlab functions for most of the procedures mentioned in this paper (MCD, LTS, MCD-regression, RQDA, ROBPCA, RPCR and RSIMPLS) are part of LIBRA, a Matlab LIBrary for Robust Analysis (Verboven and Hubert, 2005) which can be downloaded from **http://wis.kuleuven.be/stat/robust**. Several of these functions are also available in the PLS toolbox of Eigenvector Research (**www.eigenvector.com**).

## APPENDIX

### A.1 The FAST-MCD Algorithm

Rousseeuw and Van Driessen (1999) developed the FAST-MCD algorithm to efficiently compute the MCD. The key component is the C-step:

THEOREM. *Take* $\mathbf{X} = \{\mathbf{x}_1, \ldots, \mathbf{x}_n\}$ *and let* $H_1 \subset \{1, \ldots, n\}$ *be an h-subset, that is,* $|H_1| = h$. *Put* $\hat{\boldsymbol{\mu}}_1 := \frac{1}{h} \sum_{i \in H_1} \mathbf{x}_i$ *and* $\hat{\boldsymbol{\Sigma}}_1 := \frac{1}{h} \sum_{i \in H_1} (\mathbf{x}_i - \hat{\boldsymbol{\mu}}_1)(\mathbf{x}_i - \hat{\boldsymbol{\mu}}_1)'$. *If* $\det(\hat{\boldsymbol{\Sigma}}_1) \neq 0$, *define the relative distances*

$$d_1(i) := \sqrt{(\mathbf{x}_i - \hat{\boldsymbol{\mu}}_1)' \hat{\boldsymbol{\Sigma}}_1^{-1} (\mathbf{x}_i - \hat{\boldsymbol{\mu}}_1)}$$
$$\text{for } i = 1, \ldots, n.$$

*Now take* $H_2$ *such that* $\{d_1(i); i \in H_2\} := \{(d_1)_{1:n}, \ldots, (d_1)_{h:n}\}$ *where* $(d_1)_{1:n} \leq (d_1)_{2:n} \leq \cdots \leq (d_1)_{n:n}$ *are the ordered distances, and compute* $\hat{\boldsymbol{\mu}}_2$ *and* $\hat{\boldsymbol{\Sigma}}_2$ *based on* $H_2$. *Then*

$$\det(\hat{\boldsymbol{\Sigma}}_2) \leq \det(\hat{\boldsymbol{\Sigma}}_1)$$

*with equality if and only if* $\hat{\boldsymbol{\mu}}_2 = \hat{\boldsymbol{\mu}}_1$ *and* $\hat{\boldsymbol{\Sigma}}_2 = \hat{\boldsymbol{\Sigma}}_1$.

If $\det(\hat{\boldsymbol{\Sigma}}_1) > 0$, the C-step yields $\hat{\boldsymbol{\Sigma}}_2$ with $\det(\hat{\boldsymbol{\Sigma}}_2) \leq \det(\hat{\boldsymbol{\Sigma}}_1)$. Note that the C stands for "concentration" since $\hat{\boldsymbol{\Sigma}}_2$ is more concentrated (has a lower determinant) than $\hat{\boldsymbol{\Sigma}}_1$. The condition $\det(\hat{\boldsymbol{\Sigma}}_1) \neq 0$ in the C-step theorem is no real restriction because if $\det(\hat{\boldsymbol{\Sigma}}_1) = 0$ we already have the minimal objective value.

In the algorithm the C-step works as follows. Given $(\hat{\boldsymbol{\mu}}_{\text{old}}, \hat{\boldsymbol{\Sigma}}_{\text{old}})$:

1. compute the distances $d_{\text{old}}(i)$ for $i = 1, \ldots, n$
2. sort these distances, which yields a permutation $\pi$ for which $d_{\text{old}}(\pi(1)) \leq d_{\text{old}}(\pi(2)) \leq \cdots \leq d_{\text{old}}(\pi(n))$
3. put $H_{\text{new}} := \{\pi(1), \pi(2), \ldots, \pi(h)\}$
4. compute $\hat{\boldsymbol{\mu}}_{\text{new}} := \text{ave}(H_{\text{new}})$ and $\hat{\boldsymbol{\Sigma}}_{\text{new}} := \text{cov}(H_{\text{new}})$.

For a fixed number of dimensions $p$, the C-step takes only O($n$) time [because $H_{\text{new}}$ can be determined in O($n$) operations without fully sorting all the $d_{\text{old}}(i)$ distances].

C-steps can be iterated until $\det(\hat{\boldsymbol{\Sigma}}_{\text{new}}) = 0$ or $\det(\hat{\boldsymbol{\Sigma}}_{\text{new}}) = \det(\hat{\boldsymbol{\Sigma}}_{\text{old}})$. The sequence of determinants obtained in this way must converge in a finite number of steps because there are only finitely many $h$-subsets. However, there is no guarantee that the final value $\det(\hat{\boldsymbol{\Sigma}}_{\text{new}})$ of the iteration process is the global minimum of the MCD objective function. Therefore an approximate MCD solution can be obtained by taking many initial choices of $H_1$, applying C-steps to each and keeping the solution with lowest determinant. For more discussion on resampling algorithms, see Hawkins and Olive (2002).

To construct an initial subset $H_1$, a random $(p+1)$-subset $J$ is drawn and $\hat{\boldsymbol{\mu}}_0 := \text{ave}(J)$ and $\hat{\boldsymbol{\Sigma}}_0 := \text{cov}(J)$ are computed. [If $\det(\hat{\boldsymbol{\Sigma}}_0) = 0$, then $J$ can be extended by adding observations until $\det(\hat{\boldsymbol{\Sigma}}_0) > 0$.] Then, for $i = 1, \ldots, n$ the distances $d_0^2(i) := (\mathbf{x}_i - \hat{\boldsymbol{\mu}}_0)' \hat{\boldsymbol{\Sigma}}_0^{-1} (\mathbf{X}_i - \hat{\boldsymbol{\mu}}_0)$ are computed and sorted into $d_0(\pi(1)) \leq \cdots \leq d_0(\pi(n))$, which leads to $H_1 := \{\pi(1), \ldots, \pi(h)\}$. This method yields better initial subsets than drawing random $h$-subsets directly, because the probability of drawing an outlier-free subset is much higher when drawing $(p+1)$-subsets than with $h$-subsets.

The FAST-MCD algorithm contains several computational improvements. Since each C-step involves the calculation of a covariance matrix, its determinant and the corresponding distances, using fewer C-steps considerably improves the speed of the algorithm. It turns out that after two C-steps, many runs that will lead to the global minimum already have a considerably smaller determinant. Therefore, the number of C-steps is reduced by applying only two C-steps on each initial subset and selecting the 10 different subsets with lowest determinants. Only for these 10 subsets, further C-steps are taken until convergence.

This procedure is very fast for small sample sizes $n$, but when $n$ grows the computation time increases due to the $n$ distances that need to be calculated in each C-step. For large $n$ FAST-MCD uses a partitioning of the dataset, which avoids doing all the calculations in the entire data. In any case, let $\hat{\boldsymbol{\mu}}_{\text{opt}}$ and $\hat{\boldsymbol{\Sigma}}_{\text{opt}}$ denote the mean and covariance matrix of the $h$-subset with lowest covariance determinant. Then the algorithm returns

$$\hat{\boldsymbol{\mu}}_{\text{MCD}} = \hat{\boldsymbol{\mu}}_{\text{opt}} \quad \text{and} \quad \hat{\boldsymbol{\Sigma}}_{\text{MCD}} = c_{h,n} \hat{\boldsymbol{\Sigma}}_{\text{opt}},$$

where $c_{h,n}$ is the product of a consistency factor and a finite-sample correction factor (Pison, Van Aelst and Willems, 2002). Note that the FAST-MCD algorithm is itself affine equivariant.

## A.2 The FAST-LTS Algorithm

The basic component of the LTS algorithm is again the C-step, which now says that starting from an initial $h$-subset $H_1$ or an initial fit $\hat{\boldsymbol{\theta}}_1$, we can construct a new $h$-subset $H_2$ by taking the $h$ observations with smallest absolute residuals $|r_i(\hat{\boldsymbol{\theta}}_1)|$. Applying LS to $H_2$ then yields a new fit $\hat{\boldsymbol{\theta}}_2$ which is guaranteed to have a lower objective function (11).

To construct the initial $h$-subsets the algorithm starts from randomly drawn $(p+1)$-subsets. For each $(p+1)$-subset the coefficients $\boldsymbol{\theta}_0$ of the hyperplane through the points in the subset are calculated. [If a $(p+1)$-subset does not define a unique hyperplane, then it is extended by adding more observations until it does.] The corresponding initial $h$-subset is then formed by the $h$ points closest to the hyperplane (i.e., with smallest residuals). As was the case for the MCD, also here this approach yields much better initial fits than would be the case if random $h$-subsets were drawn directly.

Let $\hat{\boldsymbol{\theta}}_{\text{opt}}$ denote the least squares fit of the optimal $h$-subset found by the whole resampling procedure; then FAST-LTS returns

$$\hat{\boldsymbol{\theta}}_{\text{LTS}} = \hat{\boldsymbol{\theta}}_{\text{opt}}$$

and

$$\hat{\sigma}_{\text{LTS}} = c_{h,n} \sqrt{\frac{1}{h} \sum_{i=1}^{h} (r(\hat{\boldsymbol{\theta}}_{\text{opt}})^2)_{i:n}}.$$

## A.3 The ROBPCA Algorithm

First, the data are preprocessed by reducing their data space to the subspace spanned by the $n$ observations. This is done by a singular value decomposition of $\mathbf{X}_{n,p}$. As a result, the data are represented using at most $n - 1 = \text{rank}(\tilde{\mathbf{X}}_{n,p})$ variables without loss of information.

In the second step of the ROBPCA algorithm, a measure of outlyingness is computed for each data point. This is obtained by projecting the high-dimensional data points on many univariate directions. On every direction the univariate MCD estimator of location and scale is computed, and for every data point its standardized distance to that center is measured. Finally for each data point its largest distance over all the directions is considered. The $h$ data points with smallest outlyingness are kept, and from the covariance matrix $\mathbf{\Sigma}_h$ of this $h$-subset we select the number $k$ of principal components to retain.

The last stage of ROBPCA consists of projecting the data points onto the $k$-dimensional subspace spanned by the largest eigenvectors of $\mathbf{\Sigma}_h$ and of computing their center and shape using the weighted MCD estimator. The eigenvectors of this scatter matrix then determine the robust principal components, and the location estimate serves as a robust center.

### A.4 The SIMPLS Algorithm

The solution of the maximization problem (26) is found by taking $\mathbf{r}_1$ and $\mathbf{q}_1$ as the first left and right singular eigenvectors of $\hat{\mathbf{\Sigma}}_{xy}$. The other PLSR weight vectors $\mathbf{r}_a$ and $\mathbf{q}_a$ for $a = 2, \ldots, k$ are obtained by imposing an orthogonality constraint to the elements of the scores. If we require that $\sum_{i=1}^n t_{ia} t_{ib} = 0$ for $a \neq b$, a deflation of the cross-covariance matrix $\hat{\mathbf{\Sigma}}_{xy}$ provides the solutions for the other PLSR weight vectors. This deflation is carried out by first calculating the $x$-loading $\mathbf{p}_a = \hat{\mathbf{\Sigma}}_x \mathbf{r}_a / (\mathbf{r}_a' \hat{\mathbf{\Sigma}}_x \mathbf{r}_a)$ with $\hat{\mathbf{\Sigma}}_x$ the empirical variance–covariance matrix of the $x$-variables. Next an orthonormal base $\{\mathbf{v}_1, \ldots, \mathbf{v}_a\}$ of $\{\mathbf{p}_1, \ldots, \mathbf{p}_a\}$ is constructed and $\hat{\mathbf{\Sigma}}_{xy}$ is deflated as

$$\hat{\mathbf{\Sigma}}_{xy}^a = \hat{\mathbf{\Sigma}}_{xy}^{a-1} - \mathbf{v}_a (\mathbf{v}_a' \hat{\mathbf{\Sigma}}_{xy}^{a-1})$$

with $\hat{\mathbf{\Sigma}}_{xy}^1 = \hat{\mathbf{\Sigma}}_{xy}$. In general the PLSR weight vectors $\mathbf{r}_a$ and $\mathbf{q}_a$ are obtained as the left and right singular vectors of $\hat{\mathbf{\Sigma}}_{xy}^a$.

### ACKNOWLEDGMENTS

We would like to thank Sanne Engelen, Karlien Vanden Branden and Sabine Verboven for help with preparing the figures of this paper.